\documentclass[letterpaper,titlepage,11pt]{article}

\pdfoutput=1

\usepackage{amsmath}
\usepackage{amsfonts}
\usepackage{amssymb}
\usepackage{graphicx} 
\usepackage{subfigure}
\usepackage{enumerate}
\usepackage[
      colorlinks=true,
      linkcolor=blue,
      urlcolor=blue,
      filecolor=black,
      citecolor=red,
      pdfstartview=FitV,
      pdftitle={},
        pdfauthor={Roberto Emparan, Marina Martinez},
        pdfsubject={},
        pdfkeywords={},
        pdfpagemode=None,
        bookmarksopen=true
      ]{hyperref}

\DeclareMathOperator{\sn}{sn}

\def\clock{{\count0=\time
           \divide\count0 60
           \ifnum\count0<10 0\fi\the\count0
           \multiply\count0 -60 \advance\count0 \time
           :\ifnum\count0<10 0\fi \the\count0
         }}
\newcommand{\timestamp}{{\small\vbox{\hbox{\tt\jobname.tex}
\hbox{\the\day/\the\month/\the\year, \clock}}}}

\newcommand{\ie}{{\it i.e.,\,}}
\newcommand{\eg}{{\it e.g.,\,}}

\newcommand{\lp}{\left(}
\newcommand{\rp}{\right)}
\newcommand{\lc}{\left[}
\newcommand{\rc}{\right]}

\newcommand{\mc}[1]{\mathcal{#1}}

\newcommand{\beq}{\begin{equation}}
\newcommand{\eeq}{\end{equation}}
\newcommand{\beqa}{\begin{eqnarray}}
\newcommand{\eeqa}{\end{eqnarray}}
\newcommand{\OO}{\mathcal{O}}

\numberwithin{equation}{section}
\begin{document}
\begin{titlepage}
\leftline{}
\vskip 2cm
\centerline{\LARGE \bf Exact Event Horizon of a Black Hole Merger}
\vskip 1.2cm
\centerline{\bf Roberto Emparan$^{a,b}$, Marina Mart{\'\i}nez$^{b}$,}
\vskip 0.5cm
\centerline{\sl $^{a}$Instituci\'o Catalana de Recerca i Estudis
Avan\c cats (ICREA)}
\centerline{\sl Passeig Llu\'{\i}s Companys 23, E-08010 Barcelona, Spain}
\smallskip
\centerline{\sl $^{b}$Departament de F{\'\i}sica Fonamental, Institut de
Ci\`encies del Cosmos,}
\centerline{\sl  Universitat de
Barcelona, Mart\'{\i} i Franqu\`es 1, E-08028 Barcelona, Spain}
\vskip 0.5cm

\vskip 1.2cm
\centerline{\bf Abstract} \vskip 0.3cm 
\noindent 
We argue that the event horizon of a binary black hole merger, in the extreme-mass-ratio limit where one of the black holes is much smaller than the other, can be described in an exact analytic way. This is done by tracing in the Schwarzschild geometry a congruence of null geodesics that approaches a null plane at infinity. Its form can be given explicitly in terms of elliptic functions, and we use it to analyze and illustrate the time-evolution of the horizon along the merger. We identify features such as the line of caustics at which light rays enter the horizon, and the critical point at which the horizons touch. We also compute several quantities that characterize these aspects of the merger.

\end{titlepage}
\pagestyle{empty}
\small
\normalsize
\newpage
\pagestyle{plain}
\setcounter{page}{1}

\section{Introduction}

Black hole mergers occur in Nature \cite{Abbott:2016blz}. In the theory of General Relativity they are entirely described by the vacuum equations $R_{\mu\nu}=0$, but extracting the details of the fusion of the two horizons requires in general heavy computational resources. Nevertheless, we will show that there is one instance in which the event horizon of the merger becomes so simple that it can be described in an exact analytic way. 
This is the extreme-mass-ratio (EMR) limit in which one of the black holes is much smaller than the other. If $m$ and $M$ are the two black hole masses, or equivalently their characteristic sizes (in units $G=c=1$), then the EMR limit is $m/M\to 0$. 

This limit is often taken as one where the size of the large black hole, $M$, is fixed while the small black hole is regarded as a point-like object of size $m\to 0$. Although this viewpoint is appropriate for extracting the gravitational waves emitted in the collision (with wavelengths that grow with $M$), it erases the details of phenomena that happen on the scale of $m$, such as the evolution of the event horizon as the two black holes fuse with each other. In order to resolve these smaller length scales, we must take the EMR limit keeping $m$ fixed while $M\to\infty$.\footnote{\label{mae}These two views of the EMR limit are the leading-order approximations in a matched asymptotic expansion between the near-zone, with radii $r\ll M$, and the far-zone, with $r\gg m$, which can be matched in the overlap-zone $m\ll r\ll M$ \cite{Poisson:2011nh}. We return to this issue in the conclusions.}

The techniques and ideas that we need for describing this process are elementary. Consider the last moments before the merger, when the small black hole is at a distance $\ll M$ of the large one. The equivalence principle asserts that we can always place ourselves in the rest frame of the small black hole, and that the curvature of the large black hole can be neglected over distances $\ll M$. Then the spacetime around the small black hole should be well approximated by the Schwarzschild geometry \cite{Schwarzschild:1916uq}. Although the curvature created by the large black hole vanishes in this limit, its horizon is still present: it becomes an infinite, Rindler-type, acceleration horizon. More precisely, it is a congruence of light rays that reach asymptotic null infinity as a planar null surface. We conclude that in the EMR limit, on scales much smaller than $M$, the event horizon of the black hole merger can be found by tracing an appropriate family of light rays in the Schwarzschild geometry; specifically, a congruence of null geodesics that approach a planar horizon at a large distance from the small black hole. 

We will construct this event horizon explicitly, and show that it does indeed exhibit the behavior expected of the merger: at early times, spatial sections of the event horizon consist of two components, one of them an almost spherical small black hole, and the other an almost planar large black hole. The two horizons deform each other through their gravitational attraction (which the large black hole exerts as an acceleration effect, in accord with the equivalence principle) and develop conical shapes along a line of caustics where light rays enter the horizon before the merger. When the black holes merge, they form a single smooth surface that then relaxes down to a planar horizon at late times. We illustrate this with pictures drawn using our exact results. We also compute several parameters that characterize the merger --- they are solutions of transcendental equations, so we obtain them numerically.

In our analysis the small black hole plunges head-on into the large one, but it is easy to show that if there is a relative velocity between the two black holes, \eg the small black hole moves in a direction parallel to the large horizon, the situation is equivalent to our construction up to a rotation.

While we are not aware that this analysis of the horizon of EMR mergers has been done before, related ideas have been employed in recent years. Refs.~\cite{Amsel:2007cw,Figueras:2009iu,Emparan:2013fha,Sun:2014xoa} apply the idea that the event horizon for the fall of any gravitating object into an acceleration horizon is obtained by appropriate light-ray-tracing in the spacetime of that object. Ref.~\cite{Hamerly:2010cr} studies the event horizon of the same EMR merger as we do, but it focuses on scales $\sim M$ and therefore misses the structure of the merger that we observe.
A different study of an exact merger, focusing on two equal-mass charged black holes in the Kastor-Traschen solution in deSitter space \cite{Kastor:1992nn}, reaches some conclusions that agree with ours and are presumably generic \cite{Caveny:2003pc}. 

Finally, since we have the exact geometry for the merger --- \ie the Schwarzschild metric --- it is also possible to study the evolution of its apparent horizon. We leave this for a forthcoming article \cite{ah}.

\smallskip

\textit{Note:} a 3D animation of the horizon merger is available at the arXiv website at \url{http://arxiv.org/src/1603.00712/anc}

\section{Defining the Event Horizon}

As we explained above, the exact geometry for the merger in the limit $m/M\to 0$ is the Schwarzschild solution with mass $m$.  We seek the event horizon as a particular null hypersurface in this geometry.
Conventionally, the Schwarzschild solution has an event horizon at $r=2m$, which is a cylindrical null hypersurface that reaches $\mathcal{I}^+$ at infinite retarded time. However, we are interested instead in a different null hypersurface, namely one that reaches $\mathcal{I}^+$ at a finite retarded time with the geometry of a null plane, like an acceleration horizon would do. This acceleration horizon is the limiting form of the event horizon of the large black hole when $M\to\infty$.

So we begin with the Schwarzschild black hole, in $D=n+3$ dimensions,
\beq
ds^2=-\lp 1-\frac{r_0^n}{r^n}\rp dt^2+\frac{dr^2}{1-\frac{r_0^n}{r^n}}+r^2d\Omega^2_{(n+1)}\,.
\eeq
We use the horizon radius $r_0$ instead of the mass $m\propto r_0^n$. Although we could set $r_0=1$ without loss of generality, we will mostly keep it explicit.

This geometry has a timelike Killing vector $\partial_t$, which defines the rest frame of the small black hole, and an exact $SO(n+2)$ rotational symmetry. Both isometries are only approximate when the ratio $m/M$ is finite, and would be broken by corrections in an expansion in $m/M$. But the exact symmetry in the limit $m/M\to 0$ is crucial for our analysis.

The tangent vector to the light-ray trajectories is 
\beq
P^\mu =\frac{dx^\mu(\lambda)}{d\lambda}=(\dot{t},\dot{r},\dot{\phi}_1,\dots,\dot{\phi}_{n+1})\,,
\eeq
with $P^2=0$ and $\lambda$ an affine parameter along the geodesics. The event horizon of the collision has $SO(n+1)$ symmetry along the axis that joins the two black holes,\footnote{This is also a symmetry of a head-on, radial plunge of two Schwarzschild black holes at finite $m/M$.} so we need only consider one angle of $S^{n+1}$, call it $\phi$. Specifically, we write  
\beq\label{omegaphi}
d\Omega_{(n+1)}=d\theta^2+\sin^2\theta\, d\phi^2+\cos^2\theta\, d\Omega_{(n-1)}\,,
\eeq
and study geodesics on the plane $\theta=\pi/2$. We put the collision axis along the two segments $\phi=0,\pi$. Before the merger, $\phi=0$ points away from the large black hole and $\phi=\pi$  points towards it. 

The Killing vectors $\partial_t$ and $\partial_\phi$ of the geometry imply two integrals of motion, and the equations to solve are
\beqa
\dot{t}&=&\frac{1}{1-r_0^n/r^n}\,,\label{tdot}\\
\dot{\phi}&=&-\frac{q}{r^2}\,,\label{phidot}\\
\dot{r}&=&\frac{1}{r}\sqrt{r^2-q^2\lp 1-\frac{r_0^n}{r^n}\rp}\,,\label{rdot}
\eeqa
where $q$ is the impact parameter, \ie the ratio between the conserved angular momentum and the energy of the light-ray trajectory.

It will be convenient to use $r$ instead of $\lambda$ as the (non-affine) parameter along the geodesics. This is because the integration of \eqref{rdot} gives $\lambda(r)$ as a combination of elliptic integrals of different kinds, which we cannot invert analytically to find $r(\lambda)$ and then obtain $t(\lambda)$ and $\phi(\lambda)$. Instead, we get the geodesics as
\beq\label{ints}
t_q(r)=\int dr\, \frac{\dot{t}}{\dot{r}}\,,\qquad \phi_q(r)=\int dr\,\frac{\dot{\phi}}{\dot{r}}\,.
\eeq

The section of the event horizon in the space $(t,\phi,r)$ is a two-dimensional surface, \ie a one-parameter family of geodesics. The entire $(n+2)$-dimensional event horizon is obtained by rotating through an angle $\pi$ around the collision axis, and acting with the group $SO(n)$ to generate the $\Omega_{(n-1)}$ factor of the geometry. 

\begin{figure}
\includegraphics[scale=.75]{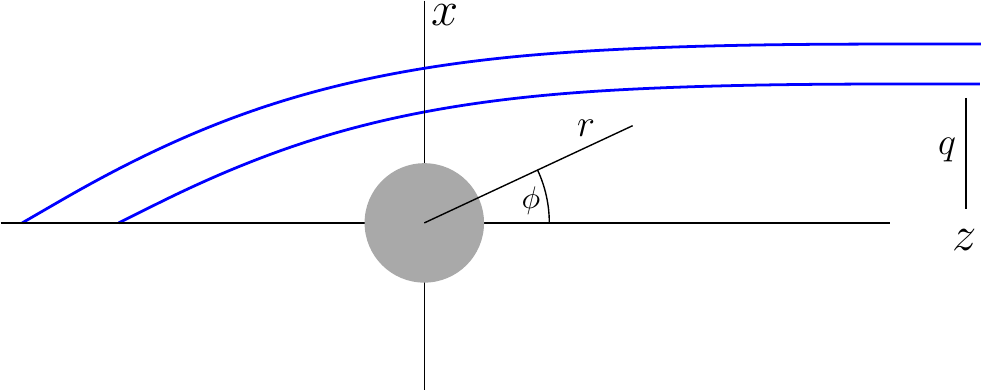} 
\centering
\caption{\small Projection on the spatial plane $(x,z)$ of null generators of the event horizon. The blue curves are the paths traced by light rays that move from left to right towards $\mathcal{I}^+$. At late times they move along the $z$ direction as the generators of a Rindler horizon ($dt=dz$). They are labelled by the impact parameter $q$ at future infinity.}
\label{qmeaning}
\end{figure}

The integration constants in \eqref{ints} are fixed by the requirement that the null surface becomes a planar horizon at infinity. The geodesics on this event horizon will be labeled by $q$.  
Let us first fix the integration constant for $\phi_q$. We have
\beq
\phi_q(r\rightarrow\infty)=\int dr\,\frac{\dot{\phi}}{\dot{r}}\bigg|_{r\rightarrow\infty}=\alpha_q+\frac{q}{r}+\mathcal{O}(r^{-3}),
\eeq
with constant $\alpha_q$. The latter corresponds to the asymptotic angle of the light-ray trajectories. In order that these rays asymptotically move all in the same direction, we must set $\alpha_q$ to a $q$-independent value. Without loss of generality, we choose 
\beq\label{alphaq}
\alpha_q=0\,.
\eeq
If we define coordinates 
\beq\label{defxz}
x=r\sin\phi\,,\qquad z=r\cos\phi\,,
\eeq
then asymptotically all light rays move with $dx=0$,
\beqa
x|_{r\rightarrow\infty}&= &q+\OO(r^{-(n+2)}),\\
z|_{r\rightarrow\infty}&= &r+\OO(r^{-1}).
\eeqa
Fig.~\ref{qmeaning} illustrates the meaning of $q$ as the impact parameter of each geodesic at infinity.

With our choice \eqref{alphaq}
the horizon will satisfy
\beq\label{dtdz}
dt-dz= \OO(r^{-n})\,.
\eeq
We fix the integration constant for $t_q$ so that all light rays arrive at $\mathcal{I}^+$ at the same, $q$-independent, retarded time. Since
\begin{align}
&t_q(r\rightarrow\infty)=r+r_0\ln\lp r/r_0\rp +\beta_q+\OO(r^{-1}), & (D=4),\\
&t_q(r\rightarrow\infty)=r+\beta_q+\OO(r^{-n}), & (D\geq 5),
\end{align}
we must set the integration constant $\beta_q$ to a $q$-independent value. For simplicity we choose
\beq\label{betaq}
\beta_q=0\,.
\eeq

The integrals \eqref{ints} do not take any simple form in general, but for $n=1,2$ they can be expressed as combinations of incomplete elliptic integrals. 
One particular generator can be found easily: the `central' geodesic at $q=0$, which is
\beqa
t_{q=0}(r)&=&\begin{cases}
r+r_0\ln\frac{r-r_0}{r_0}\,,\quad  &(n=1)\label{tq0}\\
r+r_0\ln\sqrt{\frac{r-r_0}{r+r_0}}\,,\quad  &(n=2)
\end{cases}\\
\phi_{q=0}(r)&=&0\,.
\eeqa
(the result for arbitrary $n$ can be given in terms of hypergeometric functions). This is a light ray that at $t\to -\infty$ emerges in the radial direction from the Schwarzschild horizon, to escape towards infinity.

\section{Event horizon in $\boldsymbol{D=4}$}
\begin{figure}[t]
\centering
\subfigure[]{\includegraphics[width=.49\textwidth]{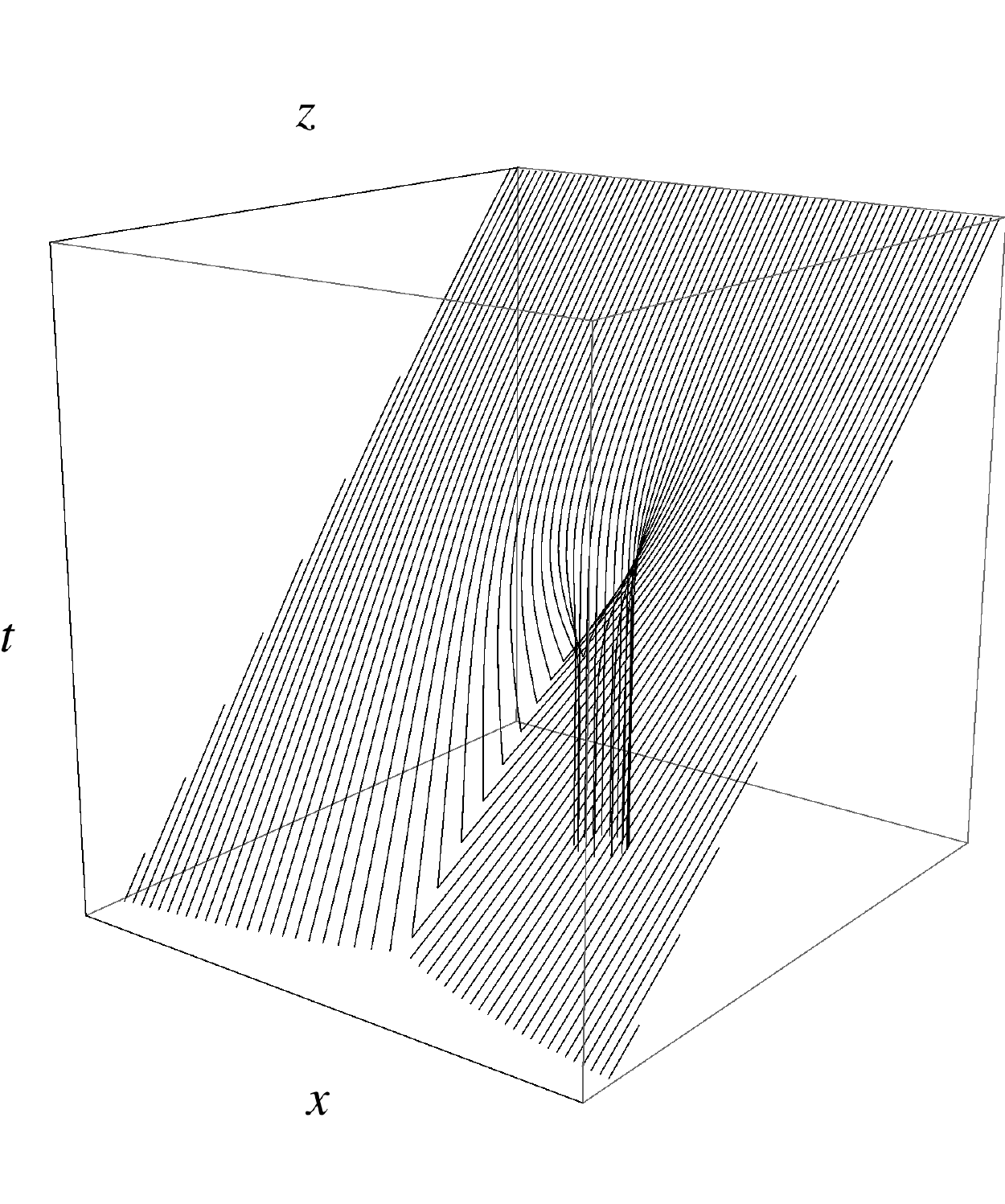}}
\subfigure[]{\includegraphics[width=.49\textwidth]{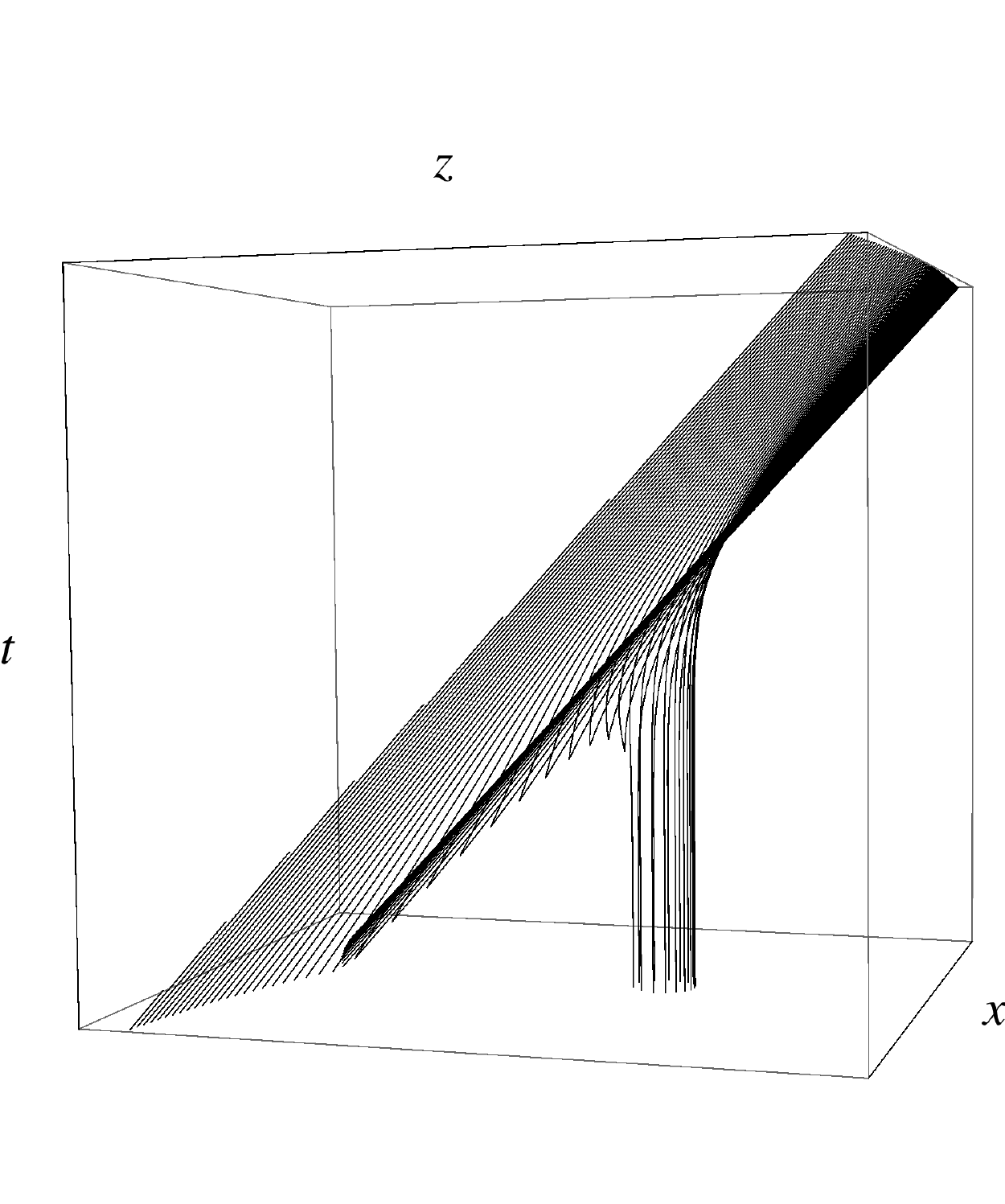}}
\caption{\small Two views of the event horizon of the four-dimensional merger, in the rest frame of the small black hole. Each curve is a null generator of the hypersurface with a different value of $q$. The coordinate $t$ is the Killing time. $\sqrt{x^2+z^2}$ is the area-radius of the Schwarzschild solution.} \label{4dh}
\end{figure}

The explicit form of the integrals for $D=4$,
\beqa
t_q(r)&=&\int\frac{r^3dr}{(r-r_0)\sqrt{r(r^3-q^2r+q^2r_0)}}\,,\label{tint}\\
\phi_q(r)&=&-\int\frac{q\ dr}{\sqrt{r(r^3-q^2r+q^2r_0)}}\,\label{phiint},
\eeqa
is not very enlightening; we give it in appendix~\ref{app:sol}. The most delicate step is fixing the integration constants to the values \eqref{alphaq} and \eqref{betaq}. We have performed first the indefinite integrals using \textsl{Mathematica}, which makes specific $q$-dependent choices for the integration constants that we must extract and then subtract. The procedure is cumbersome but straightforward. Figure \ref{4dh} shows the hypersurface generated by these geodesics.

\begin{figure}[htbp]
\centering
\subfigure[$t-t_*=-20\hspace{1pt}r_0$]{\includegraphics[width=.48\textwidth]{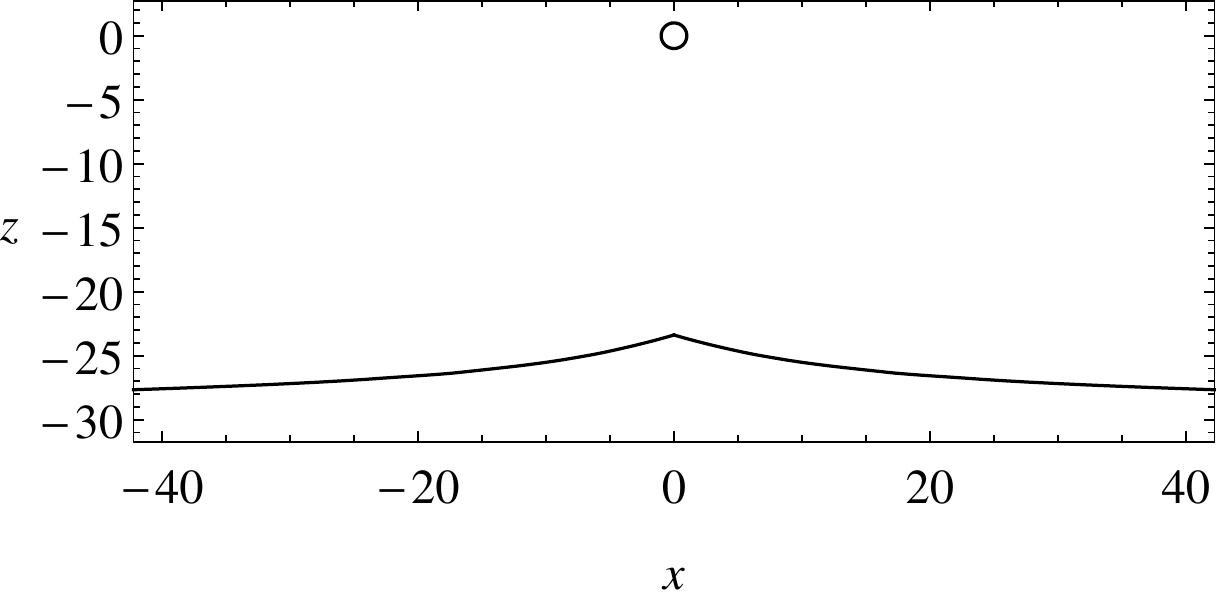}}
\subfigure[$t-t_*=-10\hspace{1pt}r_0$]{\includegraphics[width=.48\textwidth]{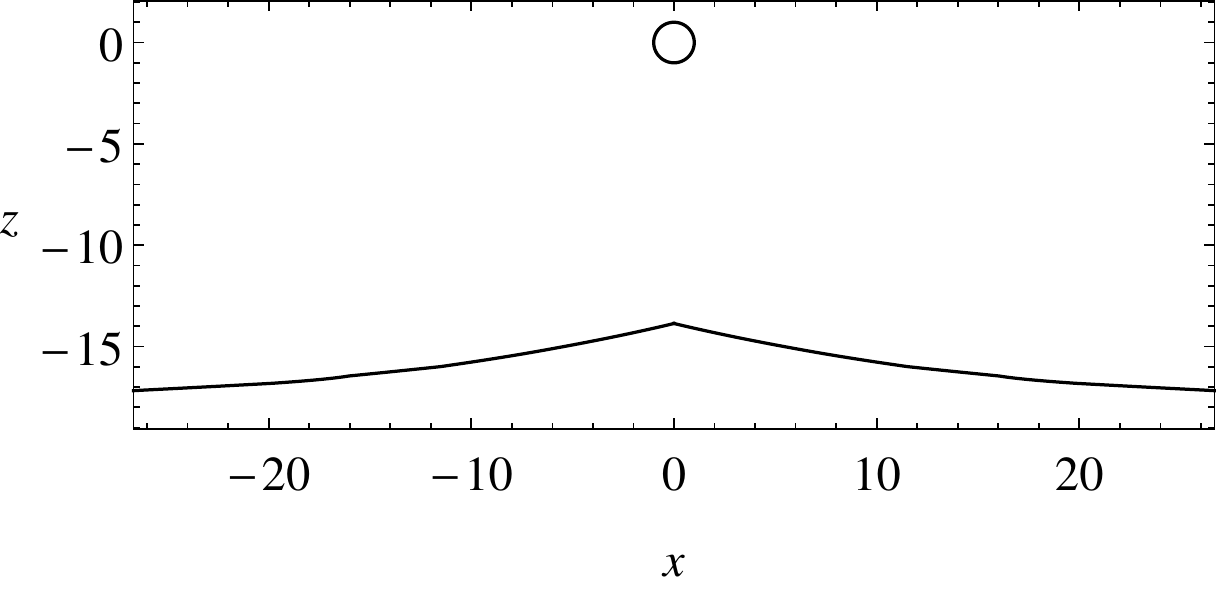}}
\subfigure[$t-t_*=-2\hspace{1pt}r_0$]{\includegraphics[width=.48\textwidth]{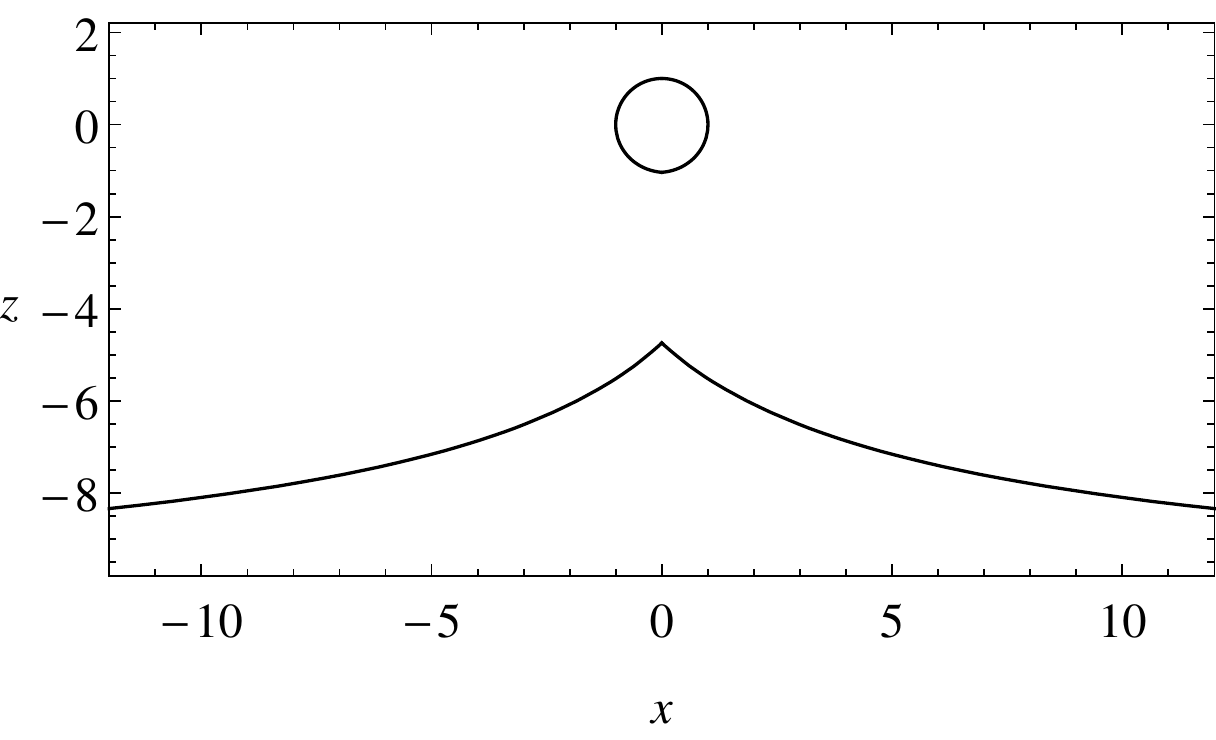}}
\subfigure[$t-t_*=-0.1\hspace{1pt}r_0$]{\includegraphics[width=.48\textwidth]{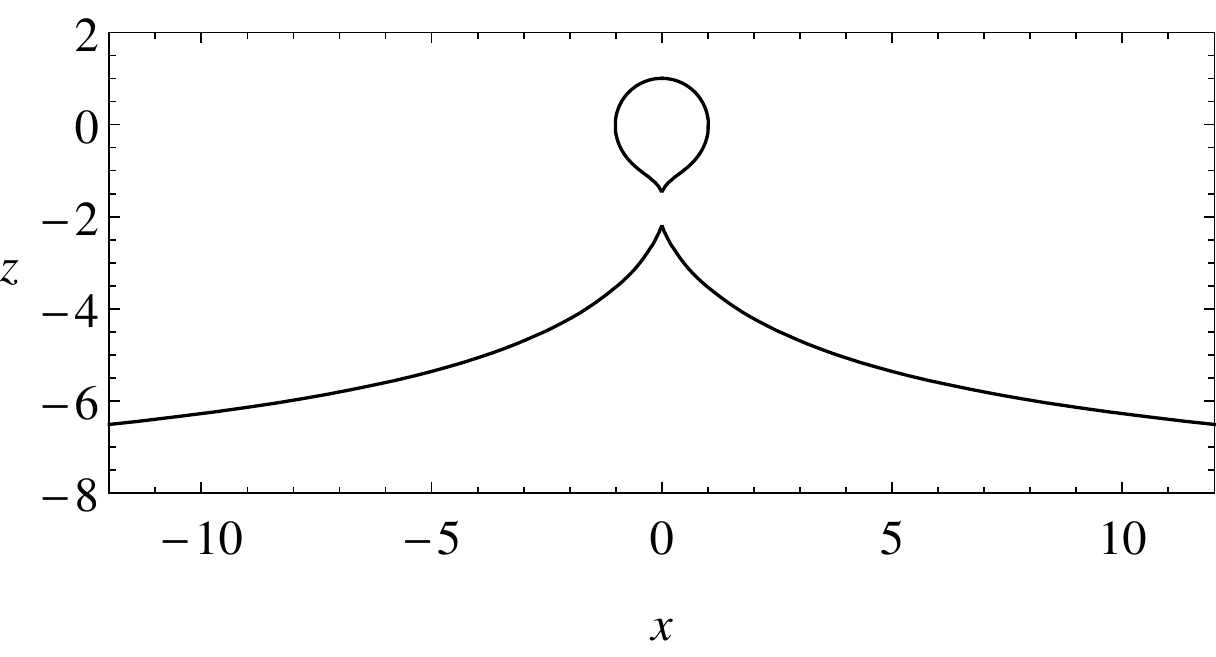}}
\subfigure[Pinch-on: $t-t_*=0$]{\includegraphics[width=.48\textwidth]{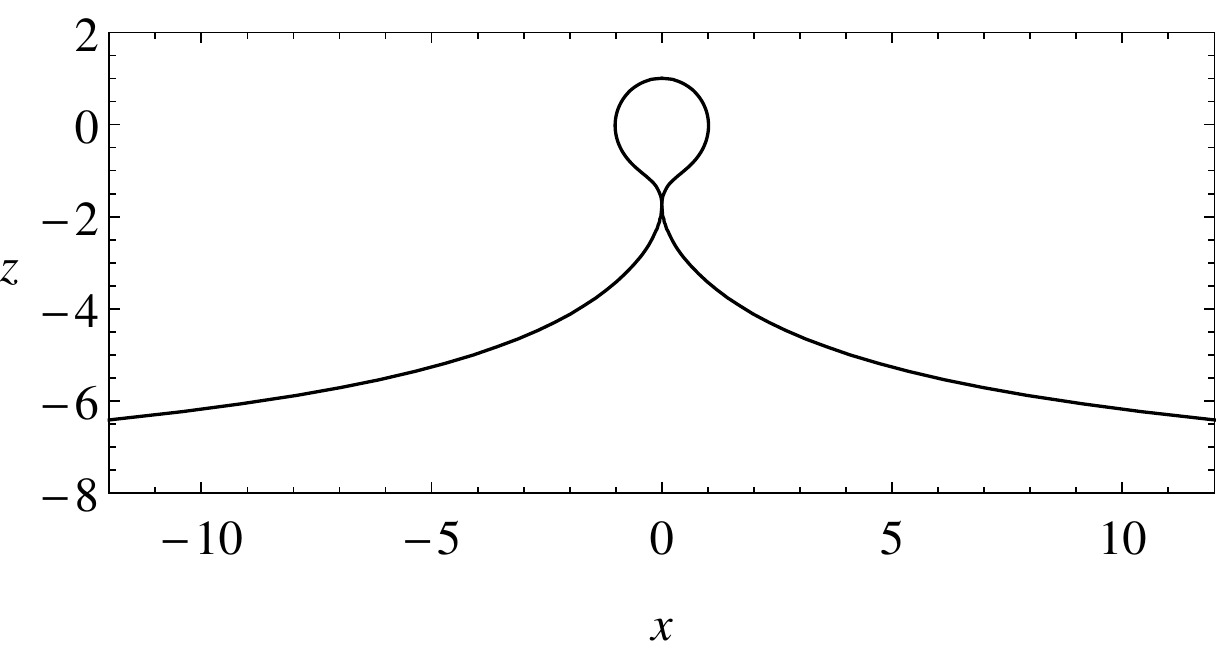}}
\subfigure[$t-t_*=\hspace{1pt}r_0$]{\includegraphics[width=.48\textwidth]{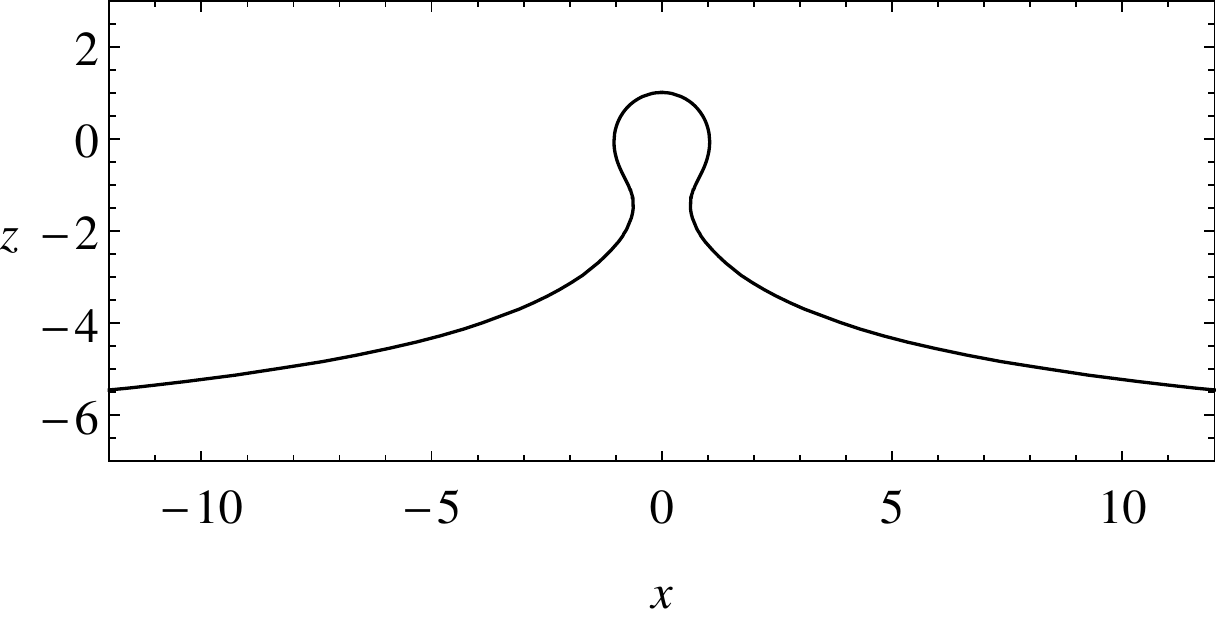}}
\subfigure[$t-t_*=\Delta_*=5.95\hspace{1pt}r_0$]{\includegraphics[width=.48\textwidth]{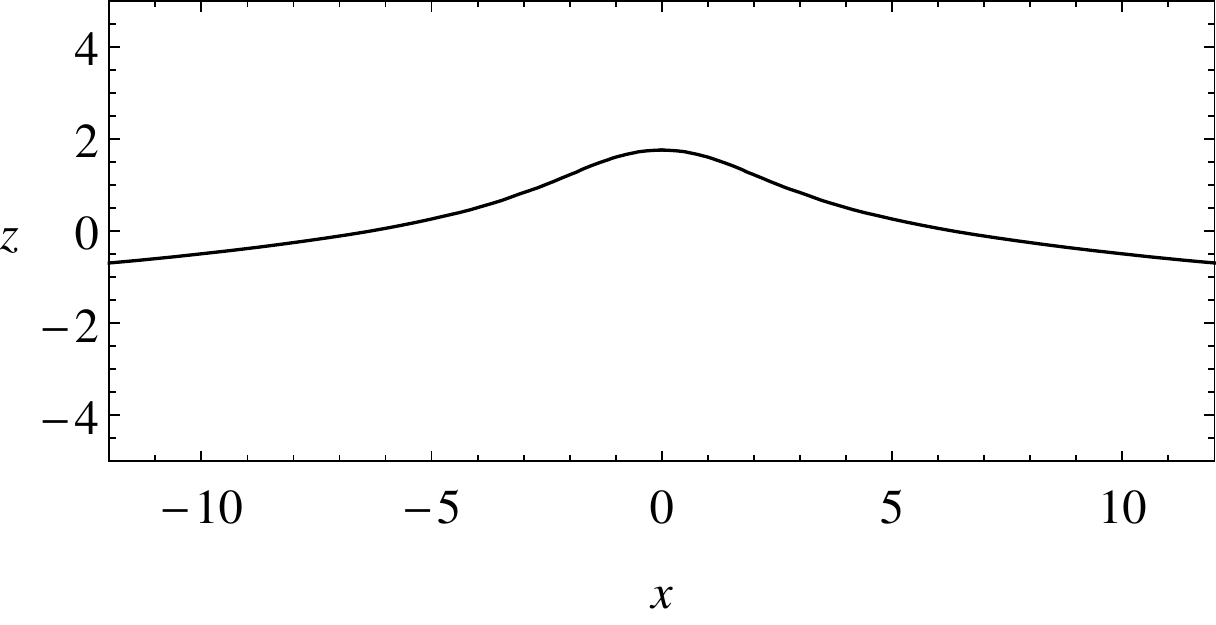}}
\subfigure[$t-t_*=27\hspace{1pt}r_0$]{\includegraphics[width=.48\textwidth]{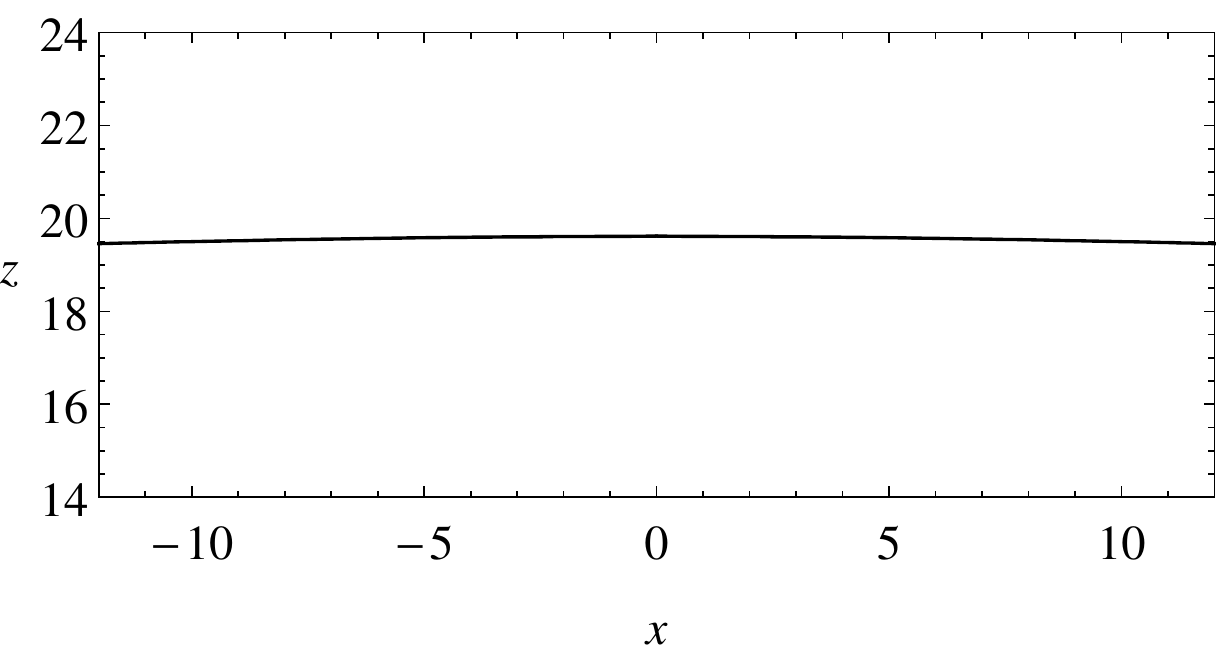}}
\caption{\small Sequence of constant-time slices of the $D=4$ event horizon. $t$ is the Killing time of the Schwarszchild geometry, and the spatial coordinates are centered on the small black hole, with area-radius $\sqrt{x^2+z^2}$.  
Pinch-on occurs at $t=t_*$ \eqref{t*}. The time interval $\Delta_*$ is a natural measure of the duration of the fusion \eqref{D*}. 
New null generators enter the two components of the horizon at all $t\in (-\infty,t_*]$, creating cones at the caustic points. The full two-dimensional constant-time slices of the event horizon are obtained by rotating around $x=0$. Axis units are $r_0=1$.} \label{cuts}
\end{figure}

In figure \ref{cuts} we show a sequence of constant-time slices of this event horizon. They clearly show the evolution expected in this merger: in the past there are two disconnected surfaces: an almost planar one and an almost spherical one. Evolving towards the future they approach each other, and eventually merge into a single horizon, which then relaxes into a flat surface. 
We do not have explicit analytic expressions for these constant-$t$ slices. These require inverting the function $t_q(r)$ to find $r_q(t)$, which we have not managed to do except in the limit $r\gg r_0$, to be discussed in sec.~\ref{subsec:pertr0}. The plots in fig.~\ref{cuts} have been drawn by taking constant-$t$ cuts of plots generated with a sufficiently dense number of geodesics. 

We often use the spatial coordinates $x$ and $z$ introduced in \eqref{defxz}. Although these are not convenient for writing the metric at finite $r$, they provide an easy way of representing the information in the plane of polar coordinates $(r,\phi)$.

\subsection{Structure of the event horizon and parameters of the merger}

Fig.~\ref{4dh} exhibits clearly the presence before the merger of a line of caustics (also known as a crease set), where light rays intersect. At these points, null generators enter to be part of the event horizon. In the full three-dimensional event horizon all the generators that intersect lie on a $S^1$ of radius $q$ at future infinity.\footnote{It may be more appropriate to refer to the intersection as a focus rather than a caustic, but the latter terminology is rather common.} The presence of caustics is generic in the event horizons of black hole mergers. In our hypersurface the caustic line extends to past infinity.

There are two special values of the impact parameter, $q_c$ and $q_*$, with $q_c<q_*$, which separate the generators into different classes.

\begin{figure}[t]
\includegraphics[scale=0.8]{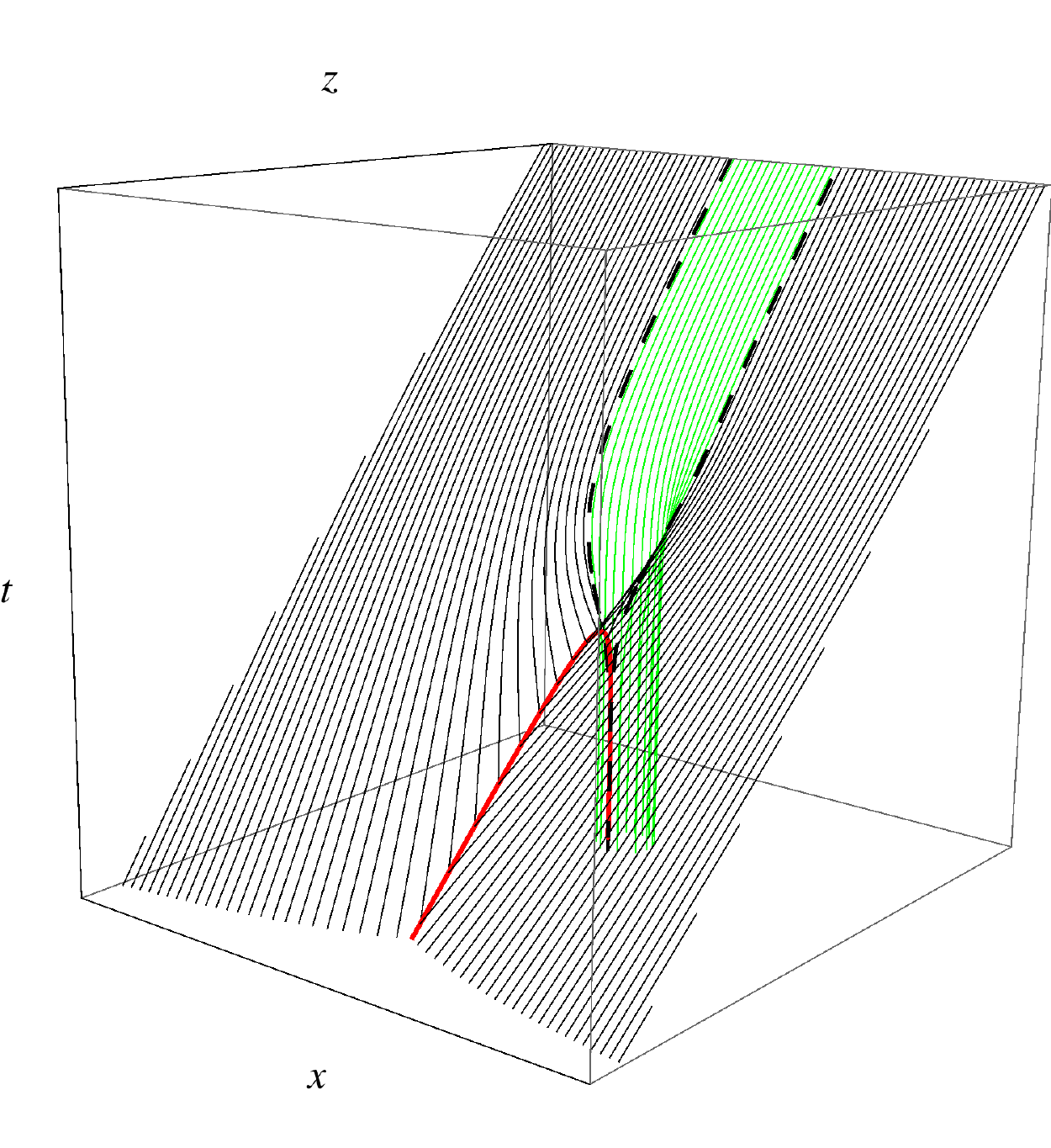} 
\centering
\caption{\small Event horizon of the merger in four dimensions. Non-caustic geodesics are the green curves, which emanate from the Schwarzschild horizon in the infinite past. Caustic geodesics are shown in black. They enter the hypersurface through the caustic line (red thick curve). The black dashed curves are the geodesics with $q=q_c$ that separate the two classes.}
\label{4dehC}
\end{figure}

\paragraph{Non-caustic generators.}
The light rays at $q=q_c$ separate the generators with $q>q_c$ that enter the horizon at a caustic at finite time, from those with $q<q_c$ that extend back to infinitely early times (see fig.~\ref{4dehC}). The latter asymptote in the past to the generators of the Schwarzschild horizon at $r=r_0$. In particular, the critical value $q=q_c$ corresponds to the rays that start at $r=r_0$ at $\phi=\pi$, and therefore are determined by the equation
\beq\label{defqc}
\phi_{q_c}(r_0)=\pi\,.
\eeq
We can solve this numerically to obtain\footnote{This and subsequent numerical solutions of transcendental equations are obtained using \textsl{Mathematica}'s \texttt{FindRoot}, which gives better precision than we are showing.}
\beq
q_c=2.22864\,r_0\,.
\eeq
The generators with $q\leq q_c$ form at future infinity a disk of radius $q_c$ and area $\pi q_c^2$. Their initial area at past infinity is the area of the Schwarzschild black hole, $\mc{A}_\textit{in}=4\pi r_0^2$. Thus in the evolution of this part of the event horizon, to which no new generators are added, the area increases by
\beq\label{delAnc}
\Delta \mc{A}_\text{non-caustic}=\lp \lp\frac{q_c}{2r_0}\rp^2-1\rp 4\pi r_0^2=0.24171\,\mc{A}_\textit{in}\,.
\eeq

\begin{figure}
\includegraphics[scale=.6]{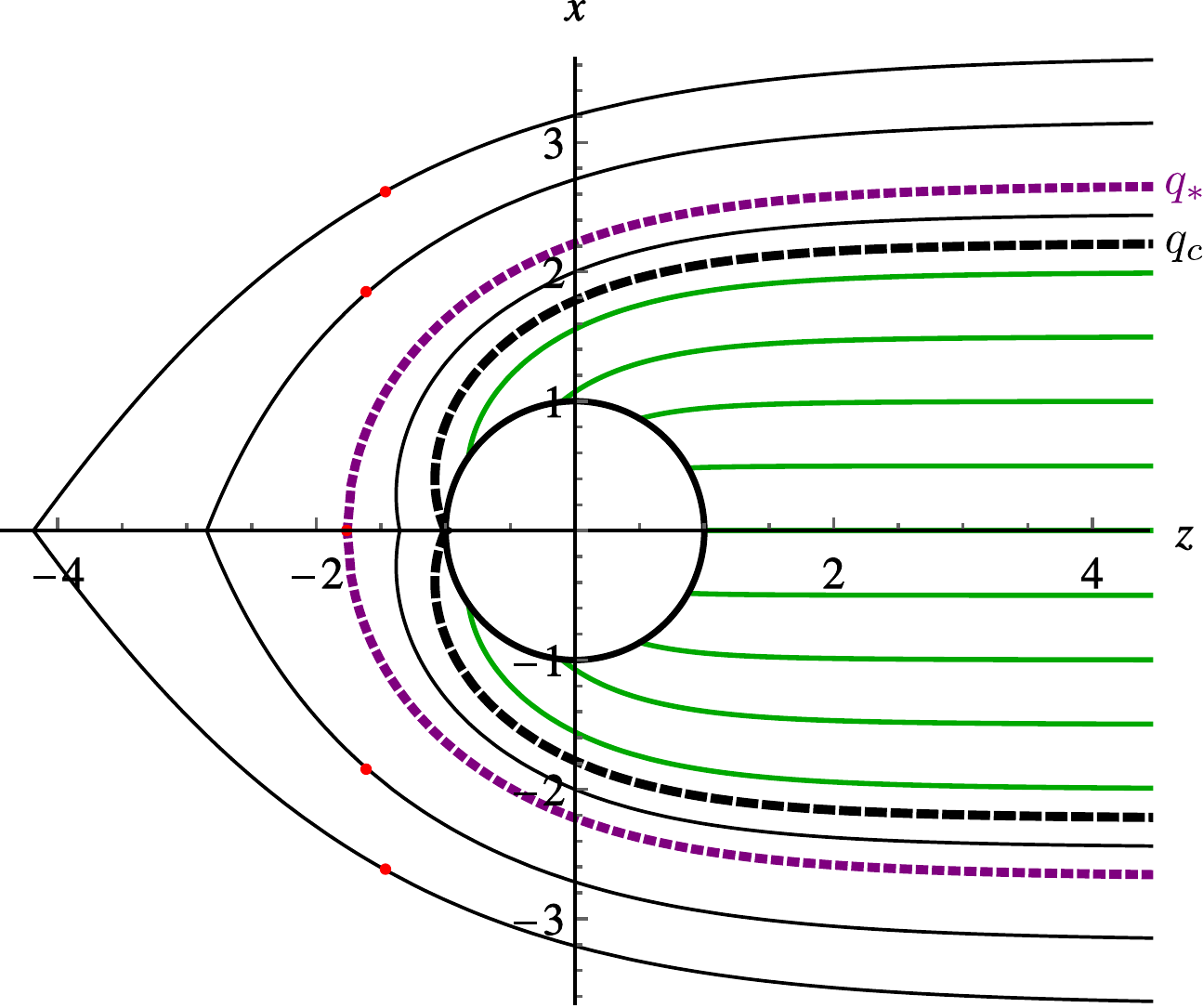} 
\centering
\caption{\small Projection of the hypersurface in the $(x,z)$ plane. All rays propagate towards $z\to+\infty$. Green curves are non-caustic generators. The black dashed curves with $q=q_c$ separate them from caustic generators. The dotted purple curves with $q=q_*$ separate the generators that enter the horizon on the side of the small black hole, $q_c<q<q_*$ from those that enter on the side of the large black hole, $q>q_*$. The latter rays first approach the small black hole, reach a minimum distance $r_\text{min}$ (red dots), and then move away from it. For $q=q_*$, the minimum $r_\text{min}$ lies at the caustic. Rays with $q_c<q<q_*$ move away from the small black hole at all times after they enter the event horizon.}
\label{xzrays}
\end{figure}

\paragraph{Caustic generators.} All generators with $q_c<q<\infty$ enter the horizon at a caustic at finite time. Among them, we single out those with $q=q_*$ which are the last to enter the horizon as measured in Killing time $t$. Generators with $q>q_*$ enter the horizon on the side of the large black hole, and generators with $q_c<q<q_*$ enter on the side of the small black hole.  This is illustrated in fig.~\ref{xzrays}.

The rays with $q=q_*$ enter at time $t=t_*$ and radius $r=r_*$. These are the parameters that characterize the pinch-on instant at which the two horizons touch and merge to form a single one.  
In order to determine these parameters, follow the rays with $q=q_*$ back in time from $\mc{I}^+$. At the caustic on the collision axis $\phi=\pi$, where they leave the horizon towards the past, they neither approach the small black hole nor escape from it, that is, $\dot{r}|_{\phi=\pi}=0$. Then \eqref{rdot} implies that $q_*$ and $r_*$ are obtained by solving the equations
\beqa
r_*^3-q_*^2 r_* +q_*^2 r_0=0\,,\qquad \phi_{q_*}(r_*)=\pi\,,
\eeqa
where $r_*$ is the largest root of the cubic polynomial.\footnote{For $q<q_\text{ph}=(3\sqrt{3}/2) r_0$ there are no real positive roots. $q_\text{ph}$ corresponds to the unstable circular photon orbit at $r_\text{ph}=3r_0/2$, which does not appear to play any special role in this construction.}  We find
\beq
q_*=2.67848\,r_0\,,\qquad r_*=1.76031\,r_0.
\eeq
$r_*$ can be taken as a measure of how strongly the small black hole is distorted, or pulled at the cusp, from the initial sphere of radius $r_0$.

Inserting these values in the solution for $t_q(r)$ we obtain
\beq\label{t*}
t_*=-4.46048\,r_0\,.
\eeq

\begin{figure}[t]
\includegraphics[scale=0.65]{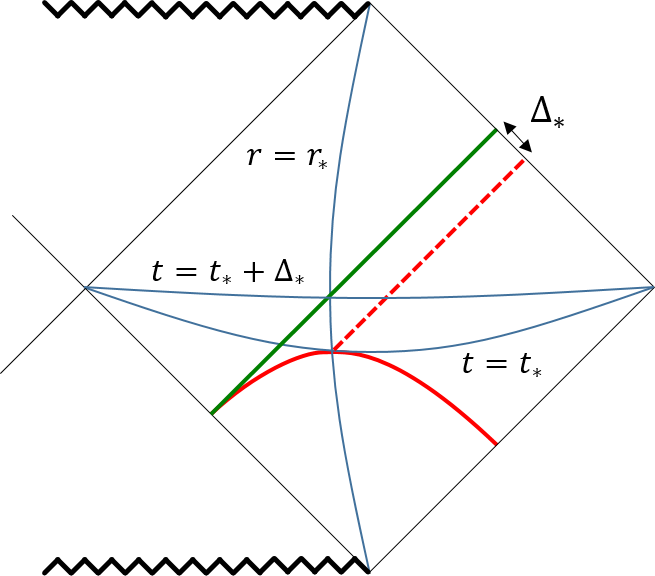} 
\centering
\caption{\small Conformal diagram for the geometry along the collision axis $\phi=0,\pi$. The green light ray is the central event horizon generator \eqref{tq0} with $q=0$, which moves along $\phi=0$. The solid-red  spacelike curve is the line of caustics, which extends along $\phi=\pi$. The dashed-red line is a light ray that emerges from the pinch-on point at $r=r_*$, $t=t_*$,  towards $\mc{I}^+$ on $\phi=\pi$. This ray propagates inside the large black hole. The retarded-time difference $\Delta_*$ characterizes the duration of the merger.}
\label{fig:cpdiagram}
\end{figure}

Note that $t_*$ is determined only with reference to the choice \eqref{betaq} that fixes the origin of retarded time. We may also consider the difference $\Delta_*$ between the retarded time at $\mc{I}^+$ of the event horizon (in the direction $\phi=0$), and the retarded time of the light ray emitted at the pinch-on instant in the opposite direction $\phi=\pi$ --- \ie\ towards the large black hole. This is
\beq\label{D*}
\Delta_*=r_* +r_0\ln((r_*-r_0)/r_0)-t_*=5.94676\,r_0\,.
\eeq
Fig.~\ref{fig:cpdiagram} illustrates it in a conformal diagram for the causal structure of the merger geometry (\ie the Schwarzschild spacetime) along the collision axis. 

Note that the dashed-red light ray propagates inside the large black hole, so the retarded-time difference $\Delta_*$ is not measurable by observers outside it.
$\Delta_*$ also admits an interpretation that does not involve any propagation through the large black hole interior: it is the time elapsed from $t_*$ until the moment when the central generator \eqref{tq0} reaches $r=r_*$ along the antipodal direction $\phi=0$, \ie\ until the instant at which the green ray intersects the line $r=r_*$ in fig.~\ref{fig:cpdiagram}. By this time, the two horizons have noticeably fused with each other (see fig.~\ref{cuts}). Through either interpretation, $\Delta_*$ can be regarded as characterizing the duration of the merger.

We can also quantify the growth in the area of the small black hole, now taking into account the addition of generators to the small horizon at caustics. This is
\beq\label{delAtot}
\Delta\mc{A}_\text{smallbh}=\lp \lp\frac{q_*}{2r_0}\rp^2-1\rp 4\pi r_0^2=0.79356\,\mc{A}_\textit{in}\,.
\eeq
The difference between \eqref{delAtot} and \eqref{delAnc} is attributed to the generators  with $q_c<q<q_*$ that are added, and to their subsequent expansion until they reach $\mc{I}^+$.

It is unclear to us whether there is any useful way to quantify the total growth of the large horizon in the merger, since it is an infinite horizon where all the generators with $q_c<q<\infty$ have entered at caustics at a finite time in the past. That is, the number of generators entering the large horizon is infinite.
This is a consequence of taking the EMR limit when the large black hole is infinite in size. Indeed, if we estimate the increment in the area neglecting the emission of radiation (see sec.~\ref{sec:conclusion}), it is expected to be
\beq
\Delta\mc{A}\simeq 16\pi (M+m)^2 -16\pi M^2 - 16\pi m^2 =32\pi M m\,,
\eeq
which diverges in the limit $M/m\to\infty$ if we keep fixed the small black hole mass $m$.

We can obtain the equation for the caustics in explicit form. In order to find the radial position $r_\text{caustic}(q)$ we have to solve
\beq
\phi_q(r_\text{caustic})=\pi
\eeq
for a given $q>q_c$.
Using the expressions in appendix~\ref{app:sol} we obtain
\beq\label{rcaustic}
r_\text{caustic}(q)=\frac{r_2+r_3}{b^{-1}\sn\lp\frac{\pi\sqrt{r_2(r_2+r_3)}}{2|q|}+F\lp\sin^{-1}\sqrt{b},\frac{r_3 a}{r_2}\rp,\frac{r_3 a}{r_2}\rp^2-1}\,,
\eeq
where $\sn$ is a Jacobi elliptic function, $r_2$ and $r_3$ are two of the roots of the polynomial $r^3-q^2r+q^2r_0$ (their explicit form is given in appendix~\ref{app:sol}), and $a$ and $b$ are combinations of them,
\begin{align}
a&=\frac{2 r_2+r_3}{r_2+2 r_3}\,,\\
b&=\frac{r_2}{2 r_2+r_3}\,.
\end{align}
Eq.~\eqref{rcaustic} is valid for both $q_c<q<q_*$ and $q>q_*$.

The rays $q_c<q<q_*$ that enter at the small horizon at radius $r=r_\text{caustic}$, afterwards move away from the small black hole in trajectories with $\dot{r}>0$. In contrast, the generators with $q>q_*$ that enter at the large horizon, first approach the small black hole, then reach a minimum distance of it, $r=r_\text{min}$, and afterwards escape away towards infinity. This minimum radius is the largest root of $r^3-q^2r+q^2r_0$ that is also smaller than $r_\text{caustic}$, and we show it for some geodesics in fig.~\ref{4dehC}. 

These considerations imply that we must be careful when computing the generators with $q>q_*$.
Since we are parametrizing the geodesics using $r$, their trajectories must be given as two branches of solutions,
\beqa
t&=&
\begin{cases}
-t_q(r)+2t_q(r_\text{min}), &r\in[r_\text{min},r_\text{caustic}], \\
t_q(r), &r\in[r_\text{min},\infty),
\end{cases}\\
\phi&=&
\begin{cases}
-\phi_q(r)+2\phi_q(r_\text{min}), &r\in[r_\text{min},r_\text{caustic}], \\
\phi_q(r), &r\in[r_\text{min},\infty), 
\end{cases}
\eeqa
where $t_q(r)$ and $\phi_q(r)$ are the same functions we use for the other geodesics.

\subsection{Perturbative solution at $r\gg r_0$}\label{subsec:pertr0}

The integrals \eqref{tint}, \eqref{phiint} simplify considerably if we evaluate them in an expansion in $r_0/r\ll 1$. In this limit the small black hole appears as a point particle, so we miss the structure of the event horizon around the region $r\sim r_0$ where the merger takes place. On the other hand, this limit allows us to find much more easily the geometry of the horizon at large distances of the small black hole. 

In fact, we can obtain explicit analytic expressions for constant-time sections of the event horizon, which we could not do in the exact solution. 
For this purpose it is convenient to give the event horizon as a surface parametrized by $t$ and $q$. To first order in $r_0$, and after fixing the integration constant \eqref{betaq} we can invert $t_q(r)$ to find
\beq
r(q,t)=\sqrt{q^2+t^2}+\frac{r_0}{2\sqrt{q^2+t^2}}\lp t-\sqrt{q^2+t^2}-2t\ln\frac{t+\sqrt{q^2+t^2}}{2r_0}\rp+\OO(r_0^2).
\eeq
With this, and after integrating $\phi_q(r)$ imposing \eqref{alphaq}, we obtain the event horizon in the spatial coordinates of \eqref{defxz},
\beqa
x(q,t)&=&q-r_0\frac{\lp t-\sqrt{q^2+t^2}\rp^2}{2q\sqrt{q^2+t^2}}+\OO(r_0^2),\notag\\
z(q,t)&=&t-\frac{r_0}{2}\lp 1-\frac{t}{\sqrt{q^2+t^2}}+2\ln\frac{t+\sqrt{q^2+t^2}}{2r_0}\rp +\OO(r_0^2)\,.\label{xzqt}
\eeqa
This is valid both for $z,t>0$ and $z,t<0$. In appendix~\ref{app:hipert} we extend it to the next order.  

The only assumption in obtaining this result is that $r\gg r_0$. 
This includes several regions of interest: 

\paragraph{Late times $t,z\gg r_0$, for all $x$ and $q$.} When $t\sim z\sim x$ (possibly $x\ll t,z$) the horizon takes an asymptotically planar form with logarithmic corrections
\beq
z=t-r_0 \lp\ln\frac{t+\sqrt{x^2+t^2}}{2r_0}+\mc{O}(1)\rp+\OO(r_0^2)\,.
\eeq

\paragraph{Moment of merger $t,z\sim r_0$, at large distance from the merger region, $|x|\gg r_0$.} Here the event horizon is the surface 
\beq
z\simeq t-r_0 \ln\frac{|x|}{2r_0}+\mc{O}(r_0^2/x)\,,
\eeq
and the spatial sections do not become flat (\ie\ $z\simeq t$) at large $|x|$, but have $z\sim -\ln |x|$ instead. 
In fact the same phenomenon happens for $t,z\gg r_0$ if we consider exponentially large $|x|\sim r_0 e^{t/r_0}$. 

This distortion from the planar shape is an effect of the long range of the gravitational field of the small black hole in four dimensions. 
In a merger of black holes with small but finite mass ratio $m/M$, the radius of the
large black hole acts as a long-distance cut-off on the coordinate $x$ along the horizon.
The results above mean that around the moment when the small black hole falls into the large one, it creates a big distortion $\sim m\ln(M/m)$ at distances $\gg m$ on the horizon of the latter, which does not dissipate until late times $t\gg m\ln(M/m)$.

\paragraph{Early times $-t,-z\gg r_0$, including the caustic line.} Here eqs.~\eqref{xzqt} apply at all $x$  and in particular around the caustic line on the large horizon at $x=0$. Then we can we study the properties of the caustic cones at early times. Assuming that the generators that reach the axis $\phi=\pi$ at this time have $q\ll |t|$ we can expand
\beqa
x(q,t)&\simeq &q-\frac{2r_0|t|}{q}\,,
\\
z(q,t)&\simeq&t-r_0\lp \ln\lp \frac{q^2}{4|t|r_0}\rp+1\rp
+\frac{2r_0^2|t|}{q^2}\,,
\eeqa
so near the axis $x=0$ we find, consistently, $q\simeq \sqrt{2r_0|t|}\ll |t|$. Note that in $z(q,t)$ we have included a term of order $r_0^2$, since near the caustic line at $|t|\gg r_0$ it contributes at the same order as the others (corrections to $z(q,t)$ at order $r_0^3$ or higher are suppressed near the caustic). 
Then, the caustic cone at the axis has slope
\beq\label{4dcone}
\left.\frac{dz}{dx}\right|_\text{cone}=\left.\frac{\partial_q z}{\partial_q x}\right|_\text{cone}\simeq -\sqrt{\frac{2r_0}{|t|}}\,.
\eeq
This result is in precise agreement with the value that \cite{Hamerly:2010cr} calculated keeping the large black hole size fixed and expanding in the small black hole size $r_0$.\footnote{Ref.~\cite{Hamerly:2010cr} in eq.~(120) gives the cone angle $\alpha$ as a function of the (large black hole) retarded time $v$ and the small black hole mass $\mu$. The result at $v\to 0$ agrees with ours at large $-t$, once we identify $2\alpha\simeq dz/dx|_\text{cone}$, $v\simeq -t$ and $\mu=r_0/2$.}

The caustic line at early times is the spacelike curve $-t=r+\mc{O}(r_0)$. Its line element is
\beq
ds_c\simeq \sqrt{\frac{r_0}{r}}dr
\eeq
so its total length $L_c=\int ds_c$ diverges in the past at $r\to\infty$. If the line is cut off by the size $\sim M$ of the large black hole, then its length is
\beq
L_c\sim \sqrt{m M}\,,
\eeq
which is in parametric agreement with the calculation in \cite{Hamerly:2010cr}.

\begin{figure}[th]
\includegraphics[scale=.85]{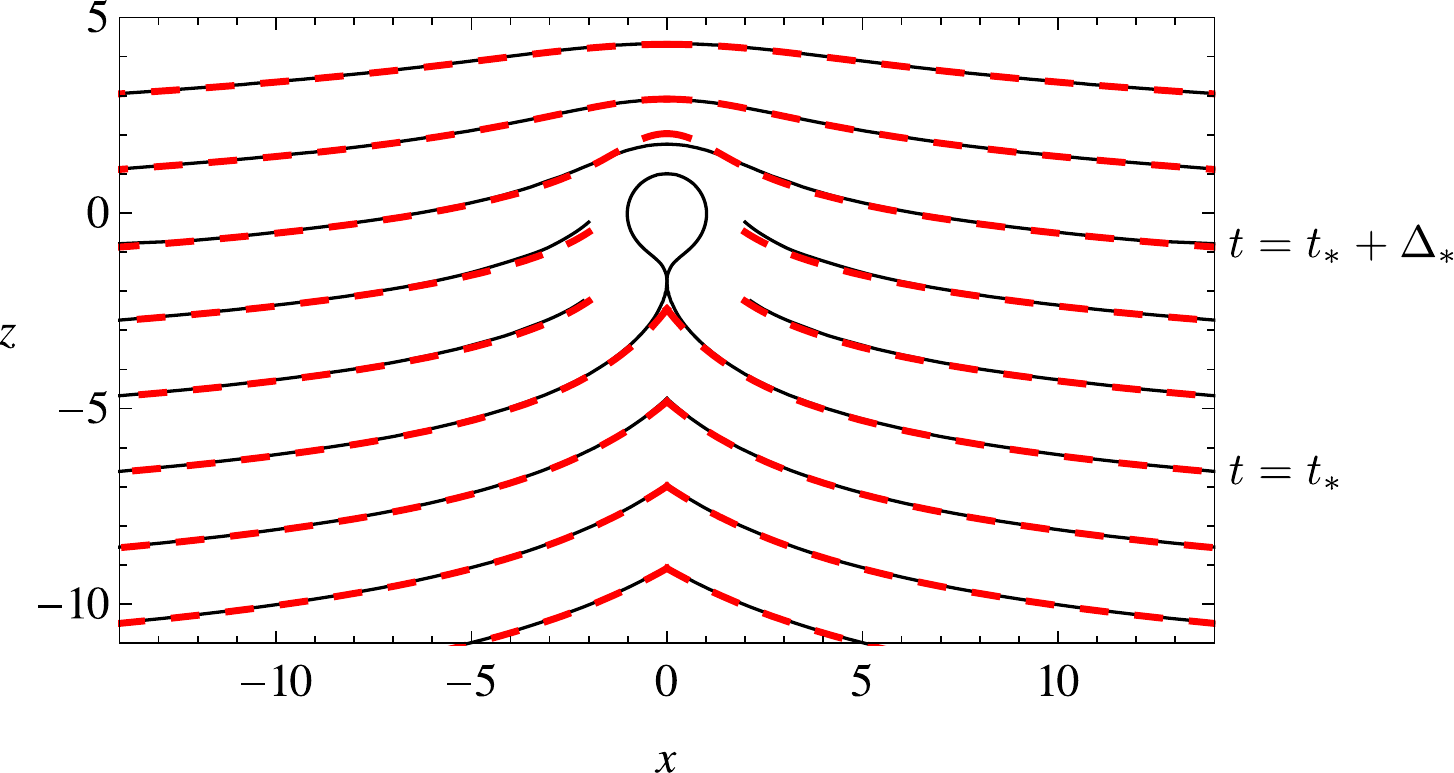} 
\centering
\caption{\small Constant-time slices of the four-dimensional event horizon, computed exactly (solid black), and approximately to order $r_0^2$ (dashed red). From top to bottom, these are $t-t_*=10,\,8,\,\Delta_*,\,4,\,2,\,0,\,-2,\,-4,\,-6$. The region excluded, where the discrepancies are larger, has diameter $\lessapprox \Delta_*$ around $x=z=0$.}
\label{fig:4dehpert}
\end{figure}

The perturbative solution at order $r_0^2$ (appendix~\ref{app:hipert}) reproduces very well the exact shape of the event horizon outside of a region of size $\approx \Delta_*$ around the merger where the evolution is very non-linear.  Fig.~\ref{fig:4dehpert} exhibits this agreement. 

It may be interesting to study the properties of the caustic line along the small black hole horizon. However, this requires a different approach and we have not pursued it.

\section{Event Horizon in $\boldsymbol{D=5}$ }

The five-dimensional version of \eqref{ints},
\beqa
t_q(r)&=&\int\frac{r^4\ dr}{(r^2-r_0^2)\sqrt{r^4-q^2r^2+q^2r_0^2}},\label{5tint}\\
\phi_q(r)&=&-\int\frac{q\ dr}{\sqrt{r^4-q^2r^2+q^2r_0^2}},\label{5phiint}
\eeqa
is not more complicated than the four-dimensional case. Once again the solution is expressed in terms of elliptic integrals of different kinds, which we give in appendix~\ref{app:sol}.

\begin{figure}[th]
\includegraphics[scale=0.7]{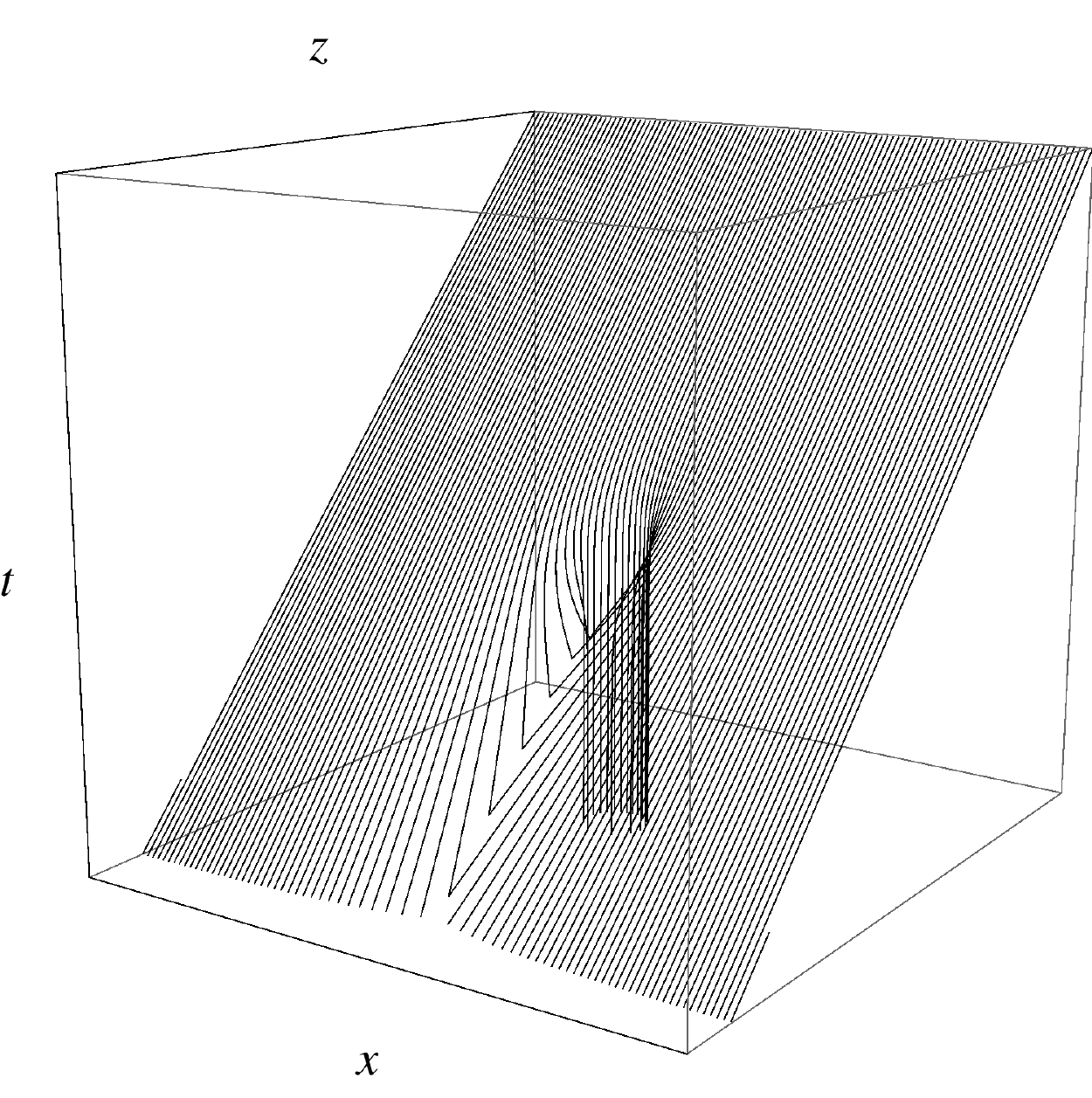} 
\centering
\caption{\small Event horizon of the merger in five dimensions.  Each curve is a null generator of the hypersurface.}
\label{5deh}
\end{figure}

We show the $D=5$ hypersurface in figure \ref{5deh}. It is very similar to the four-dimensional one, but now the event horizon at late time approaches a planar horizon more quickly, as we will see below. 

Again, we have the same three types of null geodesics on the event horizon, and the values of $q$ that delimit them are
\beq
q_c=1.88698\hspace{1pt}r_0,\qquad q_*=2.00900\hspace{1pt}r_0\,.
\eeq
The pinch-on radius and times are
\beqa
r_*&=&1.48622\,r_0\,,\\
t_*&=&-3.98444\,r_0\,,\label{t*5}\\
\Delta_*&=& 4.65473\,r_0\,.
\eeqa

All these parameters are smaller than in four dimensions, which reflects the fact that as $D$ grows larger the gravitational potential is concentrated closer to the small black hole, and the merger proceeds more swiftly.

The area increments of the small black hole part of the horizon are
\beq\label{delAnc5}
\Delta \mc{A}_\text{non-caustic}=\lp \frac{2}{3\pi}\lp\frac{q_c}{r_0}\rp^3-1\rp 2\pi^2 r_0^3=0.425807\,\mc{A}_\textit{in}\,,
\eeq
\beq\label{delAtot5}
\Delta\mc{A}_\text{smallbh}=\lp \frac{2}{3\pi}\lp\frac{q_*}{r_0}\rp^3-1\rp 2\pi^2 r_0^3=0.720674\,\mc{A}_\textit{in}\,.
\eeq
The fact that \eqref{delAnc5} is larger than in four dimensions is due to the larger number of directions in which the generators can expand. In contrast, the total increase in the area of the small black hole \eqref{delAtot5} is less than in four dimensions, indicating that less generators are added to the horizon through the milder caustic. This conforms to the general idea that black hole  mergers are less irreversible (produce less entropy) as the number of dimensions grows larger.

\paragraph{Perturbative solution.} 

Performing the integrals for small $r_0$ we now find
\beq
x(q,t)=q+r_0^2\lp -\frac{2q^2+3t^2}{4q(t^2+q^2)}+\frac{3t}{4q^2}\arctan\frac{q}{t}\rp+\OO(r_0^4)\,,
\eeq
\beq
z(q,t)=t+r_0^2\lp\frac{t}{4(t^2+q^2)}+\frac{3}{4q}\arctan\frac{q}{t}\rp+\OO(r_0^4)\,,
\eeq
when $t>0$. The analytic continuation to $t<0$ is obtained by substituting
\beq\label{arctancont}
\arctan\frac{q}{t}\to \pi -\arctan\frac{q}{|t|}\,.
\eeq
%

At late times the horizon becomes planar, $z=t+\mc{O}(r_0^2/t)$, even at very large values of $x$. It is also planar at large distances around the merger time, $|x|\gg r_0\sim t,z$, where $z=t+\mc{O}(r_0^2/x)$.

The generators that enter at the caustic at the axis $x=0$ at an early time $-t\gg r_0$ have $q^3\simeq 3\pi r_0^2|t|/4$. There the horizon develops a cone with slope
\beq\label{5dcone}
\left.\frac{dz}{dx}\right|_\text{cone}\sim -\lp\frac{r_0}{|t|}\rp^{2/3}
\eeq
(to obtain the precise factor we would need the corrections at $r_0^4$ in $z(q,t)$). The length of the caustic line $-t=r+\OO(r_0^2)$ is again infinite, but now its dependence on the large black hole size is only logarithmic,
\beq
L_c \sim \sqrt{m}\ln\frac{M}{m}\,.
\eeq

These results generalize to arbitrary dimension, where the rays at the caustic have $q^{n+1}\propto r_0^n|t|$ and
\beq\label{ncone}
\left.\frac{dz}{dx}\right|_\text{cone}\sim-\frac{q}{|t|}\sim -\lp\frac{r_0}{|t|}\rp^{n/(n+1)}\,,
\eeq
so the cone is less pointed for larger $n$.
The line element along the caustic line, $ds_c\simeq (r_0/r)^{n/2}dr$ is such that when $n>2$ the total length is finite, even if the line extends to the infinite past. This is because the caustic line approaches much more quickly a null curve, which does not add to the total proper length. 

Again, we interpret these results as consequences of the stronger localization of the gravitational field as $n$ increases, which yields a very mild caustic singularity on the horizon at $r\gg r_0$.

\section{Throat swelling}

Black hole fusion begins at the moment when the cones on the event horizon close off, develop into cusps at $t=t_*$, and then form a thin throat that connects the two horizons. We can expect that the geometry of the event horizon exhibits critical behavior in the instants before and after pinch-on. Since in this regime we do not have explicit solutions for the constant-time sections of the event horizon, we study it through slices such as those used to produce the plots in fig.~\ref{cuts}. 

Right before the merger, when the caustic cone on the large horizon closes off as $t\to t_*$, we expect that
\beq
\left.\frac{dz}{dx}\right|_\text{cone}\sim-(t_*-t)^{-\gamma}\,.
\eeq
We find $\gamma=1/2$ up to a few percent, both in $D=4$ and $D=5$. This suggests that it may be the same in other dimensions. It may be interesting to verify this, and also to investigate whether this exponent is universal, \eg\ when charge or rotation are introduced. Being a short-distance effect, the same critical behavior could be present in the small black hole horizon, but this remains to be studied.

\begin{figure}[t]
\centering
\subfigure[$\rho$ vs $(t-t_*)$ in $D=4$ merger]{\includegraphics[width=.49\textwidth]{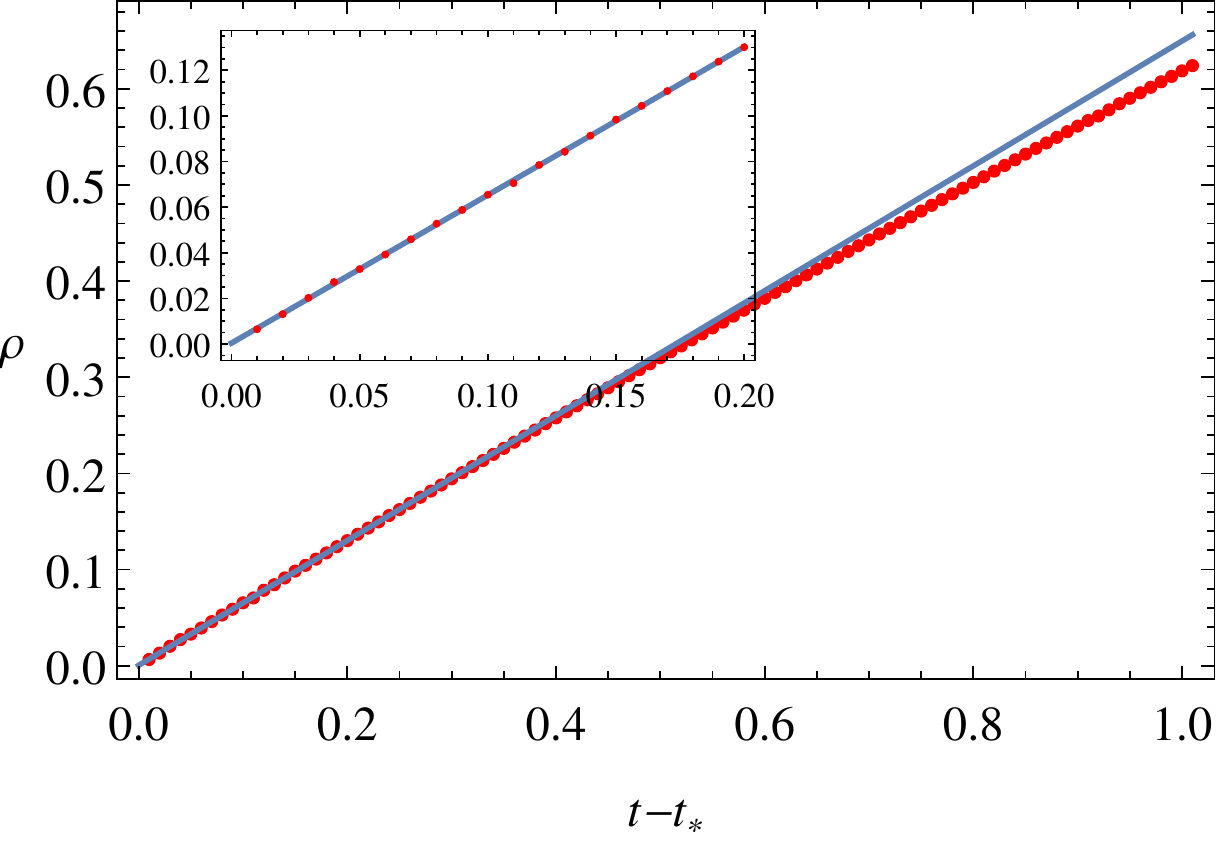}}
\subfigure[$\rho$ vs $(t-t_*)$ in $D=5$ merger]{\includegraphics[width=.49\textwidth]{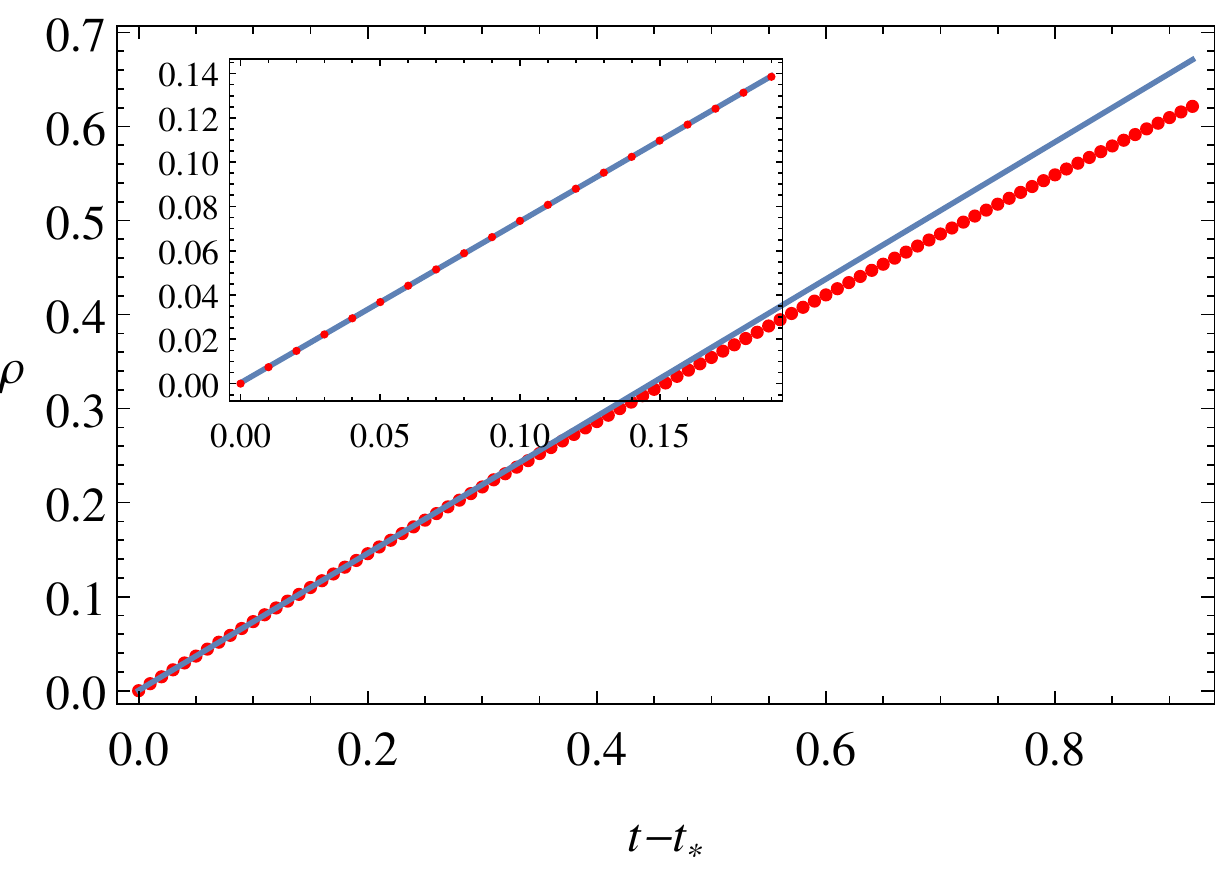}}
\caption{\small Plots of the throat thickness vs time, in units $r_0=1$. The blue curve is a linear fit to the first 20 points with $t-t_*\in (0.01,0.2)$. The inset is a magnification close to the pinch-on, where the throat grows linearly with $t$.} \label{fig:crit}
\end{figure}

We can also examine the growth of the thin throat connecting the two horizons immediately after its formation at pinch-on time $t_*$. We have managed to accurately measure the half-width $\rho$ of the throat, \ie\ its extent along the semi-axis $x>0$.
We find that, for times shortly after $t_*$, the throat grows linearly
\beq\label{linrhot}
\rho=\begin{cases}
(0.650\pm 0.005)(t-t_*)\,,\qquad (D=4)\\
(0.730\pm 0.005)(t-t_*)\,,\qquad (D=5)
\end{cases}
\eeq
(see figure \ref{fig:crit}). This linear growth lasts until $\rho/r_0\simeq 0.25$, and then begins to slow down.

The linear behavior \eqref{linrhot} agrees with the results in \cite{Caveny:2003pc}, which studied a merger in the Kastor-Traschen solution of four-dimensional charged black holes in deSitter \cite{Kastor:1992nn}. As ref.~\cite{Caveny:2003pc} observed, the linear growth may be expected since after the merger the surface is smooth. So we expect this to be a general feature of black hole mergers.

In the example in \cite{Caveny:2003pc} the behavior \eqref{linrhot} was followed after a time by exponential growth, presumably due to the deSitter expansion. Instead, in our setup the growth slows down.

\section{Mergers with relative velocities}
We have described a merger in which an infinitely large black hole approaches along the $z$ direction a finite-size black hole that is at rest. 
The asymptotic surface $dt=dz$ from which the event horizon is traced back is invariant under boosts in $z$, so the event horizon would be the same if there were any velocity along the collision axis. 

We may also consider situations where the black holes have a relative velocity along a direction parallel to the large horizon --- this includes in particular the possibility of a large rotating black hole and a small black hole on a trajectory not co-rotating with it. The generic arguments in the introduction indicate that in the EMR limit this event horizon should be equivalent to the one for radial plunge, but it may be worth elaborating the case.

Let us focus on the form of the event horizon in the asymptotic future, where we impose the initial (actually final) conditions that determine the null congruence. We have been considering that this horizon does not move, relative to the small black hole, along its planar directions, so asymptotically the surface is given by
\beq
dt=dz\,.
\label{dtdz2}
\eeq
If instead the large horizon moves along the $x$ direction, we can obtain its asymptotic form by a boost 
\beqa
t&=&\bar{t}\cosh\eta-\bar{x}\sinh\eta,\\
x&=&\bar{x}\cosh\eta-\bar{t}\sinh\eta,
\eeqa
and $z=\bar{z}$. That is, if now $\bar{t}$ is the time in the rest frame of the small black hole, the event horizon is a null congruence in the Schwarzschild geometry that asymptotes to
\beq
d\bar{t}=\frac{d\bar{z}}{\cosh\eta}+\tanh\eta\ d\bar{x}\,.
\label{boosteddtdz}
\eeq
However, this surface can also be obtained  from \eqref{dtdz2} by performing a rotation in the $(x,z)$ plane
\beqa
z&=&\bar{z}\cos\alpha+\bar{x}\sin\alpha,\\
x&=&\bar{x}\cos\alpha-\bar{z}\sin\alpha\,,
\eeqa
and $t=\bar{t}$, so that
\beq
d\bar{t}=\cos\alpha\ d\bar{z}+\sin\alpha\ d\bar{x}.
\eeq
We just need to choose the rotation angle to be
\beq\label{alphaeta}
\sin\alpha=\tanh\eta.
\eeq
Thus the effect of a boost in a direction parallel to the large horizon is equivalent to a rotation in the plane formed by this direction and the collision axis. We do not need to compute again the null generators, since it suffices to set the $q$-independent integration constant $\alpha$ in \eqref{alphaq} to the value \eqref{alphaeta}. 

\section{Concluding remarks}\label{sec:conclusion}

Everything we have needed to obtain the event horizon has been in place for a long time: the technical ingredients are the Schwarzschild solution and its null geodesics, known 100 years ago. The concepts involved are also venerably old --- the equivalence principle, which predates General Relativity itself, and the notion of event horizon, well understood more than 50 years ago.

The construction can be extended to EMR mergers with small black holes other than Schwarzschild. When the small black hole is asymptotically flat and spherically symmetric, like the Reissner-Nordstrom solution, the extension is straightforward, up to the explicit quadratures for $t_q(r)$ and $\phi_q(r)$. Of more direct physical interest is the EMR merger with a small Kerr black hole. The lower degree of symmetry makes the problem computationally quite harder, but still much simpler than when $m/M$ is finite, since we know the exact geometry in which the event horizon must be found. The class of large black holes that can be covered is also very wide, since the geometry near a non-degenerate horizon is always Rindler space. 

Gravitational wave emission is conspicuously absent from our description of the merger. In the limit $M\to\infty$ the radiation zone is pushed infinitely far away, and the geometry acquires an exact time-translation isometry, so there cannot be any waves. Relatedly, the quasinormal oscillations of the large black hole are not visible in our analysis.  The lowest quasinormal modes have wavelengths $\sim M$, so they disappear from sight in the limit, while the  higher modes of wavelength $\sim m$ have large partial wave numbers $\ell \sim M/m$ and are localized near the circular photon orbit of the large black hole, \ie\ at a large distance, $\sim M$, from the near-zone that we focus on.

Thus the price to pay for capturing exactly the event horizon is that the main observational signature of a black hole merger is removed from the picture. What is, then, the utility of this analysis? 

First of all, and leaving aside the difficulties (even of principle) of directly observing the structure of the event horizon at very short scales, we believe that it is useful to have as simple an understanding as possible of a basic phenomenon in General Relativity ---  which furthermore can be a good approximation to events that possibly take place in Nature: given the findings of \cite{Abbott:2016blz}, it does not seem impossible that black hole binary mergers with mass ratios $\lesssim 1/30$ could be detected in ground-based observatories, even more so in space-based ones. Our construction and characterization of the event horizon can be used as a benchmark for detailed numerical calculations that attempt to capture all the features of the phenomenon down to scales $\sim m$.

Second, our study gives the near-zone solution of the merger to leading order in $m/M$. One can then match it to the far-zone construction of the EMR event horizon in \cite{Hamerly:2010cr}, to obtain the first-order corrections in $m/M$. This will make the curvature of the large black hole visible in the near-zone, as well as the effects of gravitational waves on it. Corrections computed in the near-zone then provide the boundary conditions for the next-order calculation in the far zone, and so on, iteratively in a matched asymptotic expansion. It is not inconceivable that the sensitivity of future detectors will require such higher-order calculations. Our work is only the first step in the description of their event horizons.

\section*{Acknowledgements}

We are grateful to Vitor Cardoso for bringing ref.~\cite{Hamerly:2010cr} to our attention. This work has been partially supported by FPA2013-46570-C2-2-P, AGAUR 2009-SGR-168 and CPAN CSD2007-00042 Consolider-Ingenio 2010. MM is supported by an FI Fellowship of AGAUR, Generalitat de Catalunya, 2013FI B 00840.

\appendix

\section{Explicit solution}\label{app:sol}

\subsection{$\boldsymbol{D=4}$}
 In order to solve the integrals we begin by writing them as
\beqa
t_q(r)&=&\int\frac{r^3dr}{(r-r_0)\sqrt{r(r-r_1)(r-r_2)(r-r_3)}}\,,\\
\phi_q(r)&=&-\int\frac{q\ dr}{\sqrt{r(r-r_1)(r-r_2)(r-r_3)}}\,,
\eeqa
where $r_1$, $r_2$ and $r_3$ are the roots of the polynomial $r^3-q^2r+q^2r_0$. For $|q/r_0|<3\sqrt{3}/2$ one of the roots is real ($r_1$) and the other two ($r_2$ and $r_3$) are complex conjugates:
\beqa
r_1&=&\frac{-1}{\sqrt[3]{18}}\lp f_1^{1/3}+f_2^{1/3}\rp\,,\\
r_2&=&\frac{1}{\sqrt[3]{144}}\lc\lp f_1^{1/3}+f_2^{1/3}\rp+i\sqrt{3}\lp f_1^{1/3}-f_2^{1/3}\rp\rc\,,\\
r_3&=&\frac{1}{\sqrt[3]{144}}\lc\lp f_1^{1/3}+f_2^{1/3}\rp-i\sqrt{3}\lp f_1^{1/3}-f_2^{1/3}\rp\rc\,.
\eeqa
Here
\beqa
f_1&=&9q^2r_0-q^2\sqrt{81r_0^2-12q^2}\,,\\
f_2&=&9q^2r_0+q^2\sqrt{81r_0^2-12q^2}\,,
\eeqa
are positive and real for $|q/r_0|<3\sqrt{3}/2$. When $|q/r_0|>3\sqrt{3}/2$ the polynomial has three different real roots, and it is simpler to write them as
\beqa
r_1&=&-\frac{|q|}{\sqrt{3}}\lc\cos\lp\frac{1}{3}\cos^{-1}\lp\frac{-3\sqrt{3}r_0}{2|q|}\rp\rp+\sqrt{3}\sin\lp\frac{1}{3}\cos^{-1}\lp\frac{-3\sqrt{3}r_0}{2|q|}\rp\rp\rc,\\
r_2&=&\frac{2|q|}{\sqrt{3}}\lc\cos\lp\frac{1}{3}\cos^{-1}\lp\frac{-3\sqrt{3}r_0}{2|q|}\rp\rp\rc,\\
r_3&=&\frac{|q|}{\sqrt{3}}\lc-\cos\lp\frac{1}{3}\cos^{-1}\lp\frac{-3\sqrt{3}r_0}{2|q|}\rp\rp+\sqrt{3}\sin\lp\frac{1}{3}\cos^{-1}\lp\frac{-3\sqrt{3}r_0}{2|q|}\rp\rp\rc.
\eeqa
Since they are roots of a cubic with no quadratic term they satisfy 
\beq
r_1+r_2+r_3=0\,,
\eeq
which we use to eliminate $r_1$ in favor of $r_2$ and $r_3$.

The results can be expressed in terms of incomplete elliptic integrals of the first, second and third kind,
\beqa
F(x|m)&=&\int_0^x\frac{d\theta}{\sqrt{1-m\sin^2\theta}},\\
E(x|m)&=&\int_0^x \sqrt{1-m\sin^2\theta}\ d\theta,\\
\Pi(n;x|m)&=&\int_0^x\frac{d\theta}{(1-n\sin^2\theta)\sqrt{1-m\sin^2\theta}}\,.
\eeqa
When evaluating these expressions care must be exercised with the prescription for the square root of complex numbers and with the branch cuts in the elliptic functions. Our prescriptions are those implemented in \textsl{Mathematica 10}, which we have used for these calculations. 

After using identities of elliptic integrals, and fixing the integration constants to the values \eqref{alphaq}, \eqref{betaq}, we get
\begin{equation}
\phi_q(r)=\frac{2 q \left(F\left(\sin ^{-1}\left(\sqrt{\frac{b(r+r_2+r_3)}{r}}\right)\middle|\frac{r_2}{r_3}a\right)-F\left(\sin ^{-1}\left(\sqrt{b}\right)\middle|\frac{r_2}{r_3}a\right)\right)}{\sqrt{r_2 (r_2+2 r_3)}}\,,
\label{phi}
\end{equation}
and
\begin{equation}
\begin{split}
t_q(r)=&+\frac{\sqrt{r (r-r_2) (r+r_2+r_3)}}{\sqrt{r-r_3}}+r_0\ln \left(\frac{\sqrt{(r-r_2) (r-r_3)}+\sqrt{r (r+r_2+r_3)}}{\sqrt{r (r+r_2+r_3)}-\sqrt{(r-r_2) (r-r_3)}}\right)\\
&-\frac{2 r_0^3 (r_2-r_3) \Pi \left(\frac{(r_0-r_3) a}{r_0-r_2};\sin ^{-1}\left(\sqrt{\frac{r-r_2}{(r-r_3) a}}\right)\middle|\frac{r_3 a}{r_2}\right)}{\sqrt{r_2 (r_2+2 r_3)}(r_0-r_2) (r_0-r_3)}\\
&+\frac{\left(2 r_0^2 (r_3-r_2)+r_0 r_2 (r_3-r_2)+r_2 r_3 (r_2+r_3)\right) F\left(\sin ^{-1}\left(\sqrt{\frac{r-r_2}{(r-r_3) a}}\right)\middle|\frac{r_3 a}{r_2}\right)}{(r_3-r_0)\sqrt{r_2 (2 r_3+r_2)}}\\
&-\frac{2 r_0 (r_2-r_3)}{\sqrt{r_2 (r_2+2 r_3)}} \Pi \left(\frac{r_3}{r_2};\sin ^{-1}\left(\sqrt{\frac{r-r_2}{(r-r_3) a}}\right)\middle|\frac{r_3 a}{r_2}\right)\\
&-\sqrt{r_2 (r_2+2 r_3)} E\left(\sin ^{-1}\left(\sqrt{\frac{r-r_2}{(r-r_3) a}}\right)\middle|\frac{r_3 a}{r_2}\right)-c_t(q).
\end{split}
\label{t}
\end{equation}
$c_t(q)$ is the integration constant in the time integral 
\begin{equation}
\begin{split}
c_t(q)= &\ r_3-\frac{2 r_0^3 (r_2-r_3) \Pi \left(\frac{(r_0-r_3)}{(r_0-r_2)}a;\sin ^{-1}\left(\sqrt{a^{-1}}\right)\middle|\frac{r_3}{r_2}a\right)}{ \sqrt{r_2 (2 r_3+r_2)}(r_0-r_3) (r_0-r_2)}\\
&+\frac{\left(2 r_0^2 (r_3-r_2)+r_0 r_2 (r_3-r_2)+r_3 r_2 (r_3+r_2)\right) F\left(\sin ^{-1}\left(\sqrt{a^{-1}}\right)\middle|\frac{r_3 }{r_2 }a\right)}{\sqrt{r_2 (2 r_3+r_2)}(r_3-r_0)}\\
&-\frac{2 r_0 (r_2-r_3)}{\sqrt{r_2 (2 r_3+r_2)}} \Pi \left(\frac{r_3}{r_2};\sin ^{-1}\left(\sqrt{a^{-1}}\right)\middle|\frac{r_3}{r_2}a\right)\\
&+r_0 \ln \left(\frac{2}{r_3+r_2}\right)-\sqrt{r_2 (2 r_3+r_2)} E\left(\sin ^{-1}\left(\sqrt{a^{-1}}\right)\middle|\frac{r_3 }{r_2 }a\right),
\end{split}
\label{ct}
\end{equation}
and
\begin{align}
a&=\frac{2 r_2+r_3}{r_2+2 r_3},\\
b&=\frac{r_2}{2 r_2+r_3}.
\end{align}

\subsection{$\boldsymbol{D=5}$}
The integrals for the geodesics
\beqa
t_q(r)&=&\int\frac{r^4\ dr}{(r^2-r_0^2)\sqrt{r^4-q^2r^2+q^2r_0^2}},\\
\phi_q(r)&=&-\int\frac{q\ dr}{\sqrt{r^4-q^2r^2+q^2r_0^2}}\,,
\eeqa
are very similar to the four-dimensional ones. We rewrite them using two of the four roots of the polynomial under the surds
\beq
r_1=\sqrt{\frac{q-\sqrt{q^2-4r_0^2}}{2}}\,,\qquad r_2=\sqrt{\frac{q+\sqrt{q^2-4r_0^2}}{2}},
\eeq
(the other two roots are $-r_{1,2}$). Then
\beqa
t_q(r)&=&\int\frac{r^4\ dr}{(r^2-r_0^2)\sqrt{(r^2-r_1^2)(r^2-r_2^2)}},\\
\phi_q(r)&=&-\int\frac{q\ dr}{\sqrt{(r^2-r_1^2)(r^2-r_2^2)}}\,.
\eeqa
We perform the integrals again using \textsl{Mathematica 10}. After some manipulation and fixing the integration constants we obtain
\beq
\phi_q(r)=\frac{2q\lp -F\lp\sin^{-1}\lp\sqrt{\frac{(r-r_2)(r_1+r_2)}{2r_2(r-r_1)}}\rp\middle|l_+\rp+ F\lp\sin^{-1}\lp\sqrt{\frac{r_1+r_2}{2r_2}}\rp\middle|l_+\rp\rp}{r_1+r_2}
\eeq
and
\beq
\begin{split}
t_q(r)=&\sqrt{\frac{(r-r_1)(r^2-r_2^2)}{r+r_1}}-\frac{2r_1^4}{(r_1^2-r_0^2)(r_1+r_2)}F\lp\sin^{-1}\lp\sqrt{\frac{(r+r_2)m_+}{r+r_1}}\rp\middle|l_+\rp\\
&+(r_1-r_2) E\lp\sin^{-1}\lp\sqrt{\frac{(r-r_2)m_-}{r+r_1}}\rp\middle|l_-\rp +(r_1+r_2)F\lp\sin^{-1}\lp\sqrt{\frac{(r-r_2)m_-}{r+r_1}}\rp\middle|l_-\rp \\
&-\frac{r_0^3(r_1-r_2)}{(r_0-r_1)(r_0-r_2)(r_1+r_2)}\Pi\lp\frac{r_0-r_1}{(r_0-r_2)m_+};\sin^{-1}\lp\sqrt{\frac{(r+r_2)m_+}{r+r_1}}\rp\middle|l_+\rp\\
&+\frac{r_0^3(r_1-r_2)}{(r_0+r_1)(r_0+r_2)(r_1+r_2)}\Pi\lp\frac{r_0+r_1}{(r_0+r_2)m_+};\sin^{-1}\lp\sqrt{\frac{(r+r_2)m_+}{r+r_1}}\rp\middle|l_+\rp-c_t(q)
\end{split}
\eeq
where the integration constant is given by
\beq
\begin{split}
c_t(q)=&-r_1-\frac{2r_1^4}{(r_1^2-r_0^2)(r_1+r_2)}F\lp\sin^{-1}\lp\sqrt{m_+}\rp\middle|l_+\rp\\
&+(r_1-r_2) E\lp\sin^{-1}\lp\sqrt{m_-}\rp\middle|l_-\rp +(r_1+r_2)F\lp\sin^{-1}\lp\sqrt{m_-}\rp\middle|l_-\rp \\
&-\frac{r_0^3(r_1-r_2)}{(r_0-r_1)(r_0-r_2)(r_1+r_2)}\Pi\lp\frac{r_0-r_1}{(r_0-r_2)m_+};\sin^{-1}\lp\sqrt{m_+}\rp\middle|l_+\rp\\
&+\frac{r_0^3(r_1-r_2)}{(r_0+r_1)(r_0+r_2)(r_1+r_2)}\Pi\lp\frac{r_0+r_1}{(r_0+r_2)m_+};\sin^{-1}\lp\sqrt{m_+}\rp\middle|l_+\rp,
\end{split}
\eeq
and
\beq
l_\pm=\pm\frac{4r_1r_2}{(r_1\pm r_2)^2},\qquad m_\pm=\frac{r_2\pm r_1}{2r_2}.
\eeq

\section{Perturbative solution to order $r_0^2$}\label{app:hipert}

Here we give the result of the integrals \eqref{tint}, \eqref{phiint} computed up to order $r_0^2$, with the integration constants fixed to the values \eqref{alphaq} and \eqref{betaq}:
\beq
\begin{split}
t_q(r)=&\sqrt{r^2-q^2}+\frac{r_0}{2}\lp\frac{r}{\sqrt{r^2-q^2}}-1+2\ln\frac{r+\sqrt{r^2-q^2}}{2r_0}\rp\\
&-r_0^2\lp \frac{8q^2-7r^2}{8(r^2-q^2)^{3/2}}+\frac{15}{8q}\arctan\frac{q}{\sqrt{r^2-q^2}}\rp+\OO(r_0^3)\,,
\end{split}
\eeq
\beq
\begin{split}
\phi_q(r)=&\arctan\lp\frac{q}{\sqrt{r^2-q^2}}\rp+\frac{r_0}{q}\lp \frac{q^2-2r^2}{2r\sqrt{r^2-q^2}}+1\rp \\
&+r_0^2\lp-\frac{3q^4-20q^2r^2+15r^4}{16qr^2(r^2-q^2)^{3/2}}+\frac{15}{16q^2}\arctan\frac{q}{\sqrt{r^2-q^2}}\rp+\OO(r_0^3)\,.
\end{split}
\eeq

Eliminating $r$ in favor of $q$ and $t$, and using the coordinates $x$ and $z$ defined in \eqref{defxz} we find
\beq
\begin{split}
x(q,t)=&q-r_0\frac{\lp t-\sqrt{q^2+t^2}\rp^2}{2q\sqrt{q^2+t^2}}\\
&+\frac{r_0^2}{16q^2(q^2+t^2)^2}\lp 15t(q^2+t^2)^2\arctan\frac{q}{t}
-16q(q^2+t^2)^2\ln \frac{t+\sqrt{q^2+t^2}}{2r_0}\right.\\
&\left. -q\lp6q^4+15t^4+q^2t(21t+4\sqrt{q^2+t^2})-8t\sqrt{q^2+t^2}(3q^2+2t^2)\ln \frac{t+\sqrt{q^2+t^2}}{2r_0}\rp \rp\\
&+\OO(r_0^3)\,.
\end{split}
\eeq
\beq
\begin{split}
z(q,t)&=t-\frac{r_0}{2}\lp 1-\frac{t}{\sqrt{q^2+t^2}}+2\ln\frac{t+\sqrt{q^2+t^2}}{2r_0}\rp\\
&+\frac{r_0^2}{16q^2(q^2+t^2)^2}\lp 16t^4\Biggl( -t+\sqrt{q^2+t^2}\rp+q^4(-7t+4\sqrt{q^2+t^2})\\
&\phantom{+\frac{r_0^2}{16q^2(q^2+t^2)^2}\Biggl(}
\left. +q^2t^2\lp-23t+16\sqrt{q^2+t^2}\rp+15q(q^2+t^2)^2\arctan\frac{q}{t}\right.\\
&\phantom{+\frac{r_0^2}{16q^2(q^2+t^2)^2}\Biggl(} +8q^2\sqrt{q^2+t^2}(q^2+2t^2)\ln\frac{t+\sqrt{q^2+t^2}}{2r_0}\Biggr)+\OO(r_0^3)\,.
\end{split}
\eeq

These expressions are valid for $t>0$. The analytic continuation  \eqref{arctancont} gives the results for $t<0$.


\end{document}